\documentclass[aps,pra,showpacs,superscriptaddress,floatfix,onecolumn,preprint]{revtex4-1}

\usepackage{graphicx}
\usepackage{dcolumn}
\usepackage{bm}
\usepackage{setspace}
\usepackage{psfrag}
\usepackage{subfigure}
\usepackage{amsfonts}
\usepackage{braket}
\usepackage{color}

\newcommand{\be}{\begin{equation}}
\newcommand{\ee}{\end{equation}}
\newcommand{\bea}{\begin{eqnarray}}
\newcommand{\eea}{\end{eqnarray}}

\newcommand{\bef}{\begin{figure}}
\newcommand{\enf}{\end{figure}}

\begin{document}

\title{Nonadiabatic tunneling in circularly polarized laser fields: Derivation of formulas}

\author{Ingo Barth}
%\address{Max Born Institute, Max-Born-Str. 2A, 12489, Berlin, Germany}
%\ead{barth@mbi-berlin.de}
\author{Olga Smirnova}
\affiliation{Max Born Institute, Max-Born-Str. 2A, 12489, Berlin, Germany}
%\ead{smirnova@mbi-berlin.de}
\date{\today}

\begin{abstract}
We provide detailed analysis of strong field ionization of degenerate valence $p$ orbitals by circularly polarized
 fields. Our analytical approach is conceptually equivalent to the
 Perelomov, Popov, and Terent'ev (PPT) theory and is virtually exact for short range potentials.
 After benchmarking our results against the PPT theory for $s$ orbitals, we obtain the results for $p$ orbitals.
 We also show that, as long as the dipole approximation is valid,
 both the PPT method and our results are gauge invariant, in contrast with widely used strong field
 approximation (SFA).
 Our main result, which has already been
 briefly outlined in [I. Barth and O. Smirnova, Phys. Rev. A \textbf{84}, 063415 (2011)], is that
strong field ionization preferentially removes electrons counter-rotating to the circularly polarized
laser field.  The result is illustrated using the example of Kr atom.
 Strong, up to one order of magnitude, sensitivity of strong field ionization to the sense of
 electron rotation in the initial state is one of the key signatures of non-adiabatic regime
 of strong field ionization.

 %In contrast, widely used strong field approximation (SFA) is not gauge invariant and yields inaccurate results in both gauges.
%
%
%  We test technical aspects of our derivation against the PPT results for s-orbitals.
%  and it is based on Perelomov, Popov, and Terent'ev (PPT)theory.
% Since technical aspects of our method are slightly different from the original PPT theory, we first benchmark our calculation against the PPT results for s-orbitals. We then derive results for p-orbitals.
% Finally, we show that the PPT- method is gauge invariant as opposed to widely used strong field approximation (SFA).
% Our main result briefly outlined in [I. Barth, O. Smirnova, PRA ] that
%strong field ionization preferentially removes electrons counter-rotating to the field is illustrated using the example of Kr atom.
% Strong (up to one order of magnitude)sensitivity of strong field ionization to electron rotation in the initial state is one of the signatures of non-adiabatic regime of strong field ionization.
\end{abstract}
%\noindent{\it Keywords}:  strong field ionization, tunneling, ring currents, circular dichroism
\maketitle

\section{Introduction}

The analysis of ionization in strong low-frequency laser fields is often based on adiabatic approximation.
In this approximation, ionization is treated as tunneling through a static (or quasi-static) barrier
created by the binding potential and the voltage drop due to the electric field of the laser pulse.
This picture implies that the electron does not see the oscillations of the low-frequency laser field during ionization,
i.e.\ tunneling happens ``faster'' than the oscillation of linear or the rotation of circular field.
Formally, this picture corresponds to the limit $\gamma\ll1$, where $\gamma=\sqrt{2I_p}\,\omega/\mathcal{E}$ is the
Keldysh parameter \cite{keldysh}, $I_p$ is the ionization potential, $\omega$ is the laser frequency, and $\mathcal{E}$ is the
strength of the laser field.

However, for typical experimental conditions, both for linear and  for circularly polarized fields \cite{ursi1,ursi2,corkum1,ursi3,ursi4,torres},
the Keldysh parameter is often in the non-adiabatic tunneling regime \cite{Anatomy,YudinIvanov}, i.e.\,$\gamma\sim 1$.
Therefore, the adiabatic-based interpretation of these experiments is questionable.
In particular, we have shown in Ref.\,\cite{PRA1}  that for strong field ionization  in circularly polarized laser fields,
the sense of electron rotation becomes significant already for $\gamma < 1$, i.e.\ even
when using longer-wavelength laser radiation than the standard 800\,nm, e.g.\ 1300\,nm as in Ref.\,\cite{torres}.
As a consequence of non-adiabatic effects, the counter-rotating electron can have
up to one order of magnitude larger ionization rate
than co-rotating, depending on the laser field parameters.
Our theoretical prediction has now been confirmed by the experiment \cite{herath}.

%The preference for  counter-rotating electrons in strong field ionization is in contrast to the  preference for co-rotating electron  in one-photon ionization/transition from bound degenerate states \cite{bethe}
%and microwave ionization of circular Rydberg states \cite{Rzazewski1,Rzazewski2}, for discussion see Ref.\,\cite{PRA1}.

The goal of this paper is to expose our calculations and discuss the approximations we have used in deriving simple formulas for ionization rates presented in Ref.\,\cite{PRA1}.
We follow the theory developed by Perelomov, Popov, and Terent'ev (PPT) \cite{PPT1,PPT2} for short range potentials and  apply it to $p$ orbitals.
Including effects of the long-range Coulomb potential in a standard way \cite{PPT3,PPTnew}
and including non-adiabatic Coulomb effects \cite{Lisa,Jivesh} do not change our conclusions.
The Stark shift of the initial state is not included in our analysis,
but it can be calculated separately \cite{ursi4,jphysb} and used to correct the
field-free ionization potential used in the present calculation.

The key advantage of the PPT approach is its gauge invariance, which is discussed below in the paper.
The violation \cite{frolov1,frolov2,DBauer} of gauge invariance in strong field approximation (SFA) leads to both technical and
conceptual problems \cite{Smirnovajmo07}. In particular, for the ionization of $p$ orbitals by strong circularly
polarized fields,  the SFA yields inconsistent results in both gauges \cite{bauer,bauer2}, which
contradict experimental measurements of either ionization yields
\cite{herath} or electron spectra (see e.g.\ Refs.\,\cite{ursi1,ursi2,corkum1,ursi3,ursi4}).
Note that in linearly polarized fields the length gauge SFA and the PPT theory yield equivalent results for short range potentials \cite{Gribakin}.
The results of length gauge SFA and the PPT are different if long range effects are taken into account, with  length gauge SFA leading to incorrect prefactor of ionization rate \cite{keldysh}. 
The deficiencies  of velocity gauge SFA are well documented \cite{frolov1,frolov2,DBauer} and are significant even for short range potentials, 
e.g.\ velocity gauge SFA predicts identical total ionization rates from  $p_+$ and $p_-$ orbitals \cite{bauer}.

Our  analysis reveals that optimal quantum trajectory, which minimizes electron action under the barrier,
corresponds to initial non-zero lateral velocity in direction opposite to the rotation of the laser field.
The weight of this trajectory is determined %by the shape of the initial orbital and
by the direction of electron current in the initial orbital and is higher for $p_-$ orbitals in case of right circular polarization of the laser field.
%We show that the ionization rates for $p_+$ and $p_-$ orbitals in circularly polarized laser fields can be different, by the ratio up to one order of magnitude for typical experimental parameters.
%Very recently, this large difference has been  observed in the strong field sequential double ionization experiment \cite{herath}.

Finally, we show that non-adiabatic dynamics of strong field ionization leads to non-trivial rotational
dynamics of the hole left in the ion. This dynamics leads to the generation of electronic ring currents in ions \cite{barthatom,barth}, and
the coherence of this dynamics can be probed with attosecond time-resolution using attosecond transient absorbtion technique demonstrated
recently in Ref.\,\cite{goul10}.

This paper is organized as follows:
In Section II, we derive analytical formulas for the ionization rates based on the PPT theory.
We first benchmark our results for $s$ orbitals against the PPT results \cite{PPT1}.
We then derive the results for  $p_0$ and $p_\pm$ orbitals.
In Section III, we apply these formulas to the strong field ionization of the Kr atom.
Section IV concludes this work.

\section{Theory}

\subsection{Ionization model in circularly polarized laser fields}

The PPT formulas for the atomic ionization rates were derived for the strong field ionization in linearly, circularly, and elliptically polarized laser fields \cite{PPT1,PPT2}.
However, for circular and elliptical polarizations, there are 	formulas only for $s$ orbitals.
In this section, we derive the analytical formula for the ionizaton rates in circularly polarized laser fields also for $p_m$ orbitals with azimuthal quantum numbers $m=0,\pm1$.
The right ($+$) or left ($-$) circularly polarized laser field is defined as
\begin{eqnarray}
\label{eq:E}
\mathbf{E}_\pm(t)&=&\mathcal{E}\left(\cos(\omega t)\,\mathbf{e}_x\pm\sin(\omega t)\,\mathbf{e}_y\right),
\end{eqnarray}
which is connected with the vector potential
\begin{eqnarray}
\label{eq:A}
\mathbf{A}_\pm(t)&=&-A_0\left(\sin(\omega t)\,\mathbf{e}_x\mp\cos(\omega t)\,\mathbf{e}_y\right)
\end{eqnarray}
by the relation $\mathbf{E}_\pm(t)=-d\mathbf{A}_\pm(t)/dt$,
where $\mathcal{E}$ is the electric field amplitude, $A_0=\mathcal{E}/\omega$ is the velocity amplitude of the electron oscillations in the laser field, and $\omega$ is the laser frequency.

We assume that the electron ionization from the valence $p$ shell is described by the time-dependent Schr\"odinger equation (TDSE) in single active electron model within dipole approximation
\begin{eqnarray}
\label{eq:TDSE}
i\,\frac{\partial}{\partial t}\,\psi_\pm(\mathbf{r},t)&=&\left[-\frac{\nabla_{\mathbf{r}}^2}{2}+V(\mathbf{r})+\mathbf{r}\cdot\mathbf{E}_\pm(t)\right]\psi_\pm(\mathbf{r},t),
\end{eqnarray}
where $V(\mathbf{r})$ is the effective potential and the atomic units are used throughout in this work.
The exact solution of this TDSE is the integral equation for the time-dependent wavefunction $\psi_\pm(\mathbf{r},t)$ starting at time $t_0$ \cite{PPT1}
\begin{eqnarray}
\label{eq:psiex}
\psi_\pm(\mathbf{r},t)&=&\int d\mathbf{r}'\,G_\pm(\mathbf{r},t,\mathbf{r}',t_0)\psi_\pm(\mathbf{r}',t_0)-i\int_{t_0}^t dt_i\int d\mathbf{r}'\,G_\pm(\mathbf{r},t,\mathbf{r}',t_i)V(\mathbf{r}')\psi_\pm(\mathbf{r}',t_i),
\end{eqnarray}
where
\begin{eqnarray}
\label{eq:green}
G_\pm(\mathbf{r},t,\mathbf{r}',t_i)&=&\frac{\theta(t-t_i)}{(2\pi)^3}\int d\mathbf{k}\,e^{i\mathbf{v}_\pm(t)\mathbf{r}-i\mathbf{v}_\pm(t_i)\mathbf{r'}}e^{-\frac{i}{2}\int_{t_i}^t\mathbf{v}_\pm(\tau)^2\,d\tau}
\end{eqnarray}
is the Green's function of the electron for motion in a circularly polarized field,
\begin{eqnarray}
\label{eq:p}
\mathbf{v}_\pm(t)&=&\mathbf{k}+\mathbf{A}_\pm(t)
\end{eqnarray}
is the instantaneous electron velocity, and $\mathbf{k}$ is the final momentum observed at the detector.
Moreover, we divide $\mathbf{k}=\mathbf{k}_\parallel+\mathbf{k}_\perp$ into two components
$\mathbf{k}_\parallel=k_x\,\mathbf{e}_x+k_y\,\mathbf{e}_y$ and $\mathbf{k}_\perp=k_z\,\mathbf{e}_z$, which are parallel ($\mathbf{k}_\parallel\parallel\mathbf{A}_\pm(t)$)
and perpendicular ($\mathbf{k}_\perp\perp\mathbf{A}_\pm(t)$) to the laser field, respectively.

The first term of Eq.\,(\ref{eq:psiex}) does not contribute to the ionzation rate, because it describes the smearing out of the initial state \cite{PPT1}.
As in PPT theory, the main approximation of this theory is the neglect of the disortion of the initial wavefunction $\psi_\pm(\mathbf{r}',t_i)$ by Stark effect prior to ionization at time $t_i$,
i.e.\ we replace the exact wavefunction $\psi_\pm(\mathbf{r}',t_i)$ on the right side of Eq.\,(\ref{eq:psiex}) by the wavefunction of the bound orbital for the free atom $\varphi_{lm}(\mathbf{r}')e^{iI_pt_i}$
with quantum numbers $l$, $m$ and ionization potential $I_p$.
Using the field-free TDSE, the term $V(\mathbf{r}')\psi_\pm(\mathbf{r}',t_i)$ is replaced by
\begin{eqnarray}
V(\mathbf{r'})\varphi_{lm}(\mathbf{r'})e^{iI_pt_i}=\frac12(\nabla_{\mathbf{r}'}^2-2I_p)\varphi_{lm}(\mathbf{r'})e^{iI_pt_i}.
\end{eqnarray}
As already described in Ref.\ \cite{PPT1} in detail, the difference between two wavefunctions $\psi_\pm(\mathbf{r}',t_i)$ and $\varphi_{lm}(\mathbf{r}')e^{iI_pt_i}$ is small for short-range potentials,
i.e.\ the potential $V(\mathbf{r})$ falls more rapidly than the effective Coulomb potential $\sim 1/\mathbf{r}$.
However, the Coulomb corrections can be introduced using standard recipes \cite{PPT3,PPTnew} involving the time-integration of the Coulomb potential along the optimal trajectory.
In this work, we use the short-range potential and will include Coulomb corrections in our future work.

Furthermore, we assume that the laser field is turned on at $t_0\rightarrow -\infty$ adiabatically. Then, using the momentum representation of the wavefunction
\begin{eqnarray}
\label{eq:phip}
\tilde\varphi_{lm}(\mathbf{k})&=&\frac{1}{(2\pi)^{3/2}}\int d\mathbf{r}\,e^{-i\mathbf{k}\mathbf{r}}\varphi_{lm}(\mathbf{r})
\end{eqnarray}
and the abbreviation
\begin{eqnarray}
\label{eq:philm}
\phi_{lm}(\mathbf{v}_\pm(t))&=&\frac12(\mathbf{v}_\pm(t)^2+2I_p)\tilde\varphi_{lm}(\mathbf{v}_\pm(t)),
\end{eqnarray}
we get the approximative electron wavefunction from Eq.\,(\ref{eq:psiex})
\begin{eqnarray}
\label{eq:psi1}
\psi_\pm(\mathbf{r},t)&=&\frac{i}{(2\pi)^{3/2}}\int_{-\infty}^t dt_i\,e^{iI_pt_i}\int d\mathbf{k}\,e^{i\mathbf{v}_\pm(t)\mathbf{r}}e^{-\frac{i}{2}\int_{t_i}^t\mathbf{v}_\pm(\tau)^2\,d\tau}\,\phi_{lm}(\mathbf{v}_\pm(t_i)).
\end{eqnarray}
For circularly polarized fields, it is advantageous to use cylindrical coordinates $(\rho,\phi,z)$ instead of Cartesian ones $(x,y,z)$ in coordinate space related by $x=\rho\cos\phi$ and $y=\rho\sin\phi$.
Similarly, we introduce cylindrical coordinates $(k_\rho,\theta,k_z)$ in momentum space with relations $k_x=k_\rho\cos\theta$ and $k_y=k_\rho\sin\theta$,
thus $k_\rho^2=k_x^2+k_y^2=k_\parallel^2$ and $k^2=k_\rho^2+k_z^2=k_\parallel^2+k_\perp^2$.
With Eqs.\,(\ref{eq:A}) and (\ref{eq:p}), two exponents in Eq.\,(\ref{eq:psi1}) are expressed as
\begin{eqnarray}
\label{eq:pr}
i\mathbf{v}_\pm(t)\mathbf{r}&=&if_\pm(k_\rho,\theta,\phi,t)\rho+ik_zz
\end{eqnarray}
and
\begin{eqnarray}
\label{eq:intp2}
-\frac{i}{2}\int_{t_i}^t\mathbf{v}_\pm(\tau)^2\,d\tau
&=&-\frac{i}{2}\left(k^2+A_0^2\right)(t-t_i)-i\mathbf{k}(\boldsymbol\xi_\pm(t)-\boldsymbol\xi_\pm(t_i)),
\end{eqnarray}
where
\begin{eqnarray}
\label{eq:f}
f_\pm(k_\rho,\theta,\phi,t)&=&k_\rho\cos(\theta-\phi)-A_0\sin(\omega t\mp\phi)
\end{eqnarray}
and
\begin{eqnarray}
\label{eq:xi}
\boldsymbol\xi_\pm(t)&=&\mathbf{E}_\pm(t)/\omega^2.
\end{eqnarray}

\subsection{Derivation of the formula for the time-averaged ionization rate}

We follow the derivation of the formula for the ionization rate in Refs.\,\cite{PPT1,PPT2} based on the PPT approach and repeat it here for clarification and only for the case of circular polarization ($\varepsilon=1$).
The time-averaged ionization rate $w_\pm(\mathcal{E},\omega)$ is equal to the time-averaged radial flux at the infinity $\rho\rightarrow\infty$, i.e.
\begin{eqnarray}
\label{eq:ionrate}
w_\pm(\mathcal{E},\omega)&=&\lim_{\rho\rightarrow\infty}\,\overline{J_\pm(\rho,t)}.
\end{eqnarray}
The radial flux $J_\pm(\rho,t)$ is evaluated as the integral of the radial component of the flux density $j_{\rho\pm}(\mathbf{r},t)$ over a cylinder of radius $\rho$ with its axis
along the propagation $z$-axis of the circularly polarized laser field, i.e.
\begin{eqnarray}
\label{eq:jflux}
J_\pm(\rho,t)&=&\rho\int_{-\infty}^\infty dz\int_0^{2\pi}d\phi\,j_{\rho\pm}(\rho,\phi,z,t),
\end{eqnarray}
where $j_{\rho\pm}(\mathbf{r},t)$ is defined as
\begin{eqnarray}
\label{eq:j}
j_{\rho\pm}(\mathbf{r},t)&=&\frac{i}{2}\left(\psi_\pm(\mathbf{r},t)\frac{\partial}{\partial\rho}\,\psi_\pm^*(\mathbf{r},t)-\psi_\pm^*(\mathbf{r},t)\frac{\partial}{\partial\rho}\,\psi_\pm(\mathbf{r},t)\right).
\end{eqnarray}
Inserting Eqs.\,(\ref{eq:psi1})--(\ref{eq:intp2}) into Eq.\,(\ref{eq:j}), we get
\begin{eqnarray}
\label{eq:j2}
j_{\rho\pm}(\mathbf{r},t)
&=&\frac{1}{2(2\pi)^3}\int d\mathbf{k}_1\int d\mathbf{k}_2\,\,e^{i(\mathbf{k}_2-\mathbf{k}_1)(\mathbf{r}-\boldsymbol\xi_\pm(t))}
(f_\pm(k_{1\rho},\theta_1,\phi,t)+f_\pm(k_{2\rho},\theta_2,\phi,t))\\\nonumber
&&\int_{-\infty}^t dt_{1i}\,e^{\frac{i}{2}\left(k_1^2+A_0^2+2I_p\right)(t-t_{1i})}F_\pm^*(\mathbf{k}_1,t_{1i})
\int_{-\infty}^t dt_{2i}\,e^{-\frac{i}{2}\left(k_2^2+A_0^2+2I_p\right)(t-t_{2i})}F_\pm(\mathbf{k}_2,t_{2i}),
\end{eqnarray}
where the function
\begin{eqnarray}
\label{eq:Fkt}
F_\pm(\mathbf{k},t)&=&\phi_{lm}(\mathbf{v}_\pm(t))e^{i\mathbf{k}\boldsymbol\xi_\pm(t)}
\end{eqnarray}
contains terms with complicated, but periodic time-dependence.
Expanding $F_\pm(\mathbf{k},t)$ into the Fourier series
\begin{eqnarray}
F_\pm(\mathbf{k},t)=\sum_{n=-\infty}^\infty F_{n\pm}(\mathbf{k},\omega)e^{-in\omega t}
\end{eqnarray}
with the Fourier coefficients
\begin{eqnarray}
\label{eq:Fouriercoeff}
F_{n\pm}(\mathbf{k},\omega)&=&\frac{1}{2\pi}\int_{-\pi}^\pi d(\omega t)\,F_\pm(\mathbf{k},t)e^{in\omega t}
\end{eqnarray}
and carrying out time-integrations in Eq.\,(\ref{eq:j2}), according to $(a\in\mathbb{R}, \delta>0)$
\begin{eqnarray}
\int_{-\infty}^t dt'\,e^{\pm ia(t-t')}&=&\lim_{\delta\rightarrow 0}\frac{\pm i}{a\pm i\delta},
\end{eqnarray}
yields the final formula for the radial component of the flux density ($\delta>0$)
\begin{eqnarray}
\label{eq:j3}
j_{\rho\pm}(\mathbf{r},t)
&=&\lim_{\delta\rightarrow 0}\frac{1}{2(2\pi)^3}\int d\mathbf{k}_1\int d\mathbf{k}_2\,\,e^{i(\mathbf{k}_2-\mathbf{k}_1)(\mathbf{r}-\boldsymbol\xi_\pm(t))}
(f_\pm(k_{1\rho},\theta_1,\phi,t)+f_\pm(k_{2\rho},\theta_2,\phi,t))\qquad\\\nonumber
&&\sum_{n_1=-\infty}^\infty F_{n_1\pm}^*(\mathbf{k}_1,\omega)\left[\frac{k_1^2}{2}+\frac{A_0^2}{2}+I_p-n_1\omega+i\delta\right]^{-1}\\\nonumber
&&\sum_{n_2=-\infty}^\infty F_{n_2\pm}(\mathbf{k}_2,\omega)\left[\frac{k_2^2}{2}+\frac{A_0^2}{2}+I_p-n_2\omega- i\delta\right]^{-1}e^{-i(n_2-n_1)\omega t}.
\end{eqnarray}
This expression is then inserted into the formula for the the radial flux $J_\pm(\rho,t)$, Eq.\,(\ref{eq:jflux}).
With
\begin{eqnarray}
i(\mathbf{k}_2-\mathbf{k}_1)\mathbf{r}&=&i\rho((k_{2x}-k_{1x})\cos\phi+(k_{2y}-k_{1y})\sin\phi)+i(k_{2z}-k_{1z})z,
\end{eqnarray}
the $z$-integration is easily carried out, i.e.
\begin{eqnarray}
\int_{-\infty}^\infty dz\,e^{i(\mathbf{k}_2-\mathbf{k}_1)\mathbf{r}}&=&2\pi\delta(k_{2z}-k_{1z})e^{i\rho((k_{2x}-k_{1x})\cos\phi+(k_{2y}-k_{1y})\sin\phi)}.
\end{eqnarray}
But the evaluation of the $\phi$-integration is challenging.
Using Eq.\,(\ref{eq:f}), the Euler's formula, and the substitution
$|\mathbf{k}_{2\parallel}-\mathbf{k}_{1\parallel}|\,\sin\phi'=(k_{2x}-k_{1x})\cos\phi+(k_{2y}-k_{1y})\sin\phi$, the $\phi$-integral is evaluated as (see Appendix 1)
\begin{eqnarray}
\label{eq:appendix1}
&&\int_0^{2\pi}d\phi\,e^{i\rho((k_{2x}-k_{1x})\cos\phi+(k_{2y}-k_{1y})\sin\phi)}(f_\pm(k_{1\rho},\theta_1,\phi,t)+f_\pm(k_{2\rho},\theta_2,\phi,t))\\\nonumber
&=&2\pi i\left(k_{2\rho}^2-k_{1\rho}^2-2k_{2\rho} A_0\sin(\omega t\mp\theta_2)+2k_{1\rho} A_0\sin(\omega t\mp\theta_1)\right)\frac{J_1\left(\rho|\mathbf{k}_{2\parallel}-\mathbf{k}_{1\parallel}|\right)}{|\mathbf{k}_{2\parallel}-\mathbf{k}_{1\parallel}|},
\end{eqnarray}
where $J_n(x)$ is the Bessel function of the first kind, cf.\ Ref.\,\cite{PPT2}.
Then, we carry out the simple integration over $k_{1z}$ and get the result for the radial flux
\begin{eqnarray}
\label{eq:jflux1}
J_\pm(\rho,t)&=&\lim_{\delta\rightarrow 0}\frac{i}{\pi}\int d\mathbf{k}_{1\parallel}\int d\mathbf{k}_{2\parallel}\,e^{-i(\mathbf{k}_{2\parallel}-\mathbf{k}_{1\parallel})\boldsymbol\xi_\pm(t)}\\\nonumber
&&\left(k_{2\rho}^2-k_{1\rho}^2-2k_{2\rho} A_0\sin(\omega t\mp\theta_2)+2k_{1\rho} A_0\sin(\omega t\mp\theta_1)\right)\frac{\rho J_1\left(\rho|\mathbf{k}_{2\parallel}-\mathbf{k}_{1\parallel}|\right)}{|\mathbf{k}_{2\parallel}-\mathbf{k}_{1\parallel}|}\\\nonumber
&&\int_{-\infty}^\infty dk_z\sum_{n_1=-\infty}^\infty F_{n_1\pm}^*((\mathbf{k}_{1\parallel},k_z),\omega)\left[k_{1\rho}^2+k_z^2+A_0^2+2I_p-2n_1\omega+i\delta\right]^{-1}\\\nonumber
&&\sum_{n_2=-\infty}^\infty F_{n_2\pm}((\mathbf{k}_{2\parallel},k_z),\omega)\left[k_{2\parallel}^2+k_z^2+A_0^2+2I_p-2n_2\omega- i\delta\right]^{-1}e^{-i(n_2-n_1)\omega t}.
\end{eqnarray}
To obtain the limit of the radial flux at the infinity $\rho\rightarrow\infty$, we apply the relation for the arbitrary function $g(\mathbf{k}_\parallel)$ (see Appendix 2)
\begin{eqnarray}
\label{eq:appendix2}
\lim_{\rho\rightarrow\infty}\int d\mathbf{k}_\parallel\,g(\mathbf{k}_\parallel)\,\frac{\rho J_1(\rho k_\parallel)}{k_\parallel}&=&2\pi\int d\mathbf{k}_\parallel\,g(\mathbf{k}_\parallel)\delta(\mathbf{k}_\parallel),
\end{eqnarray}
cf.\ Ref.\,\cite{PPT2}, carry out the integration of the radial flux over $k_{1y}$, and use Eqs.\,(\ref{eq:E}), (\ref{eq:xi}), and the substitutions $k_\pm=k_{2x}\pm k_{1x}$.
Then, the intermediate result is
\begin{eqnarray}
\label{eq:jfluxlim}
\lim_{\rho\rightarrow\infty}J_\pm(\rho,t)&=&\lim_{\delta\rightarrow 0}i\int_{-\infty}^\infty dk_-\,\delta(k_-)k_-e^{-\frac{iA_0k_-}{\omega}\,\cos(\omega t)}\\\nonumber
&&\int_{-\infty}^\infty dk_{y}\int_{-\infty}^\infty dk_z\sum_{n_1=-\infty}^\infty\sum_{n_2=-\infty}^\infty e^{-i(n_2-n_1)\omega t}\int_{-\infty}^\infty dk_+\,h_\pm(k_+)	\\\nonumber
&&\left[\frac14(k_+-k_-)^2+k_y^2+k_z^2+A_0^2+2I_p-2n_1\omega+i\delta\right]^{-1}\\\nonumber
&&\left[\frac14(k_++k_-)^2+k_y^2+k_z^2+A_0^2+2I_p-2n_2\omega- i\delta\right]^{-1},
\end{eqnarray}
where the analytical function $h(k_+)$ is defined as
\begin{eqnarray}
\nonumber
h_\pm(k_+)&=&\left(k_+-2A_0\sin(\omega t)\right)
F_{n_1\pm}^*\left(\left(\frac{k_+-k_-}{2},k_y,k_z\right),\omega\right)F_{n_2\pm}\left(\left(\frac{k_++k_-}{2},k_y,k_z\right),\omega\right).\\
\label{eq:hk}
\end{eqnarray}
It is now shown in Eqs.\,(\ref{eq:jfluxlim}) and (\ref{eq:hk}), that the ionization rate depends on the sense $(\pm)$ of circular polarization only in the function $F_\pm(\mathbf{k},t)$, Eq.\,(\ref{eq:Fkt}).
By further deep analysis, the $k_+$-integrand in Eq.\,(\ref{eq:jfluxlim}) has four poles whose locations on the complex plane and corresponding residues depend particularly on $n_1$ and $n_2$.
For $k_-=0$ and $0\neq n_1\neq n_2\neq 0$, all four residues are finite, thus the $k_-$-integral in Eq.\,(\ref{eq:jfluxlim}) would be zero due to the appearance of the factor $k_-$ in the integrand.
Therefore, the condition for the number of photons $n=n_1=n_2$ must be satisfied.
Furthermore, for $2n\omega<k_y^2+k_z^2+A_0^2+2I_p$, there are only two residues that could contribute to the ionization rate, but in the limit $k_-=0$ these residues are opposite.
Therefore, we consider only the case $2n\omega\geq k_y^2+k_z^2+A_0^2+2I_p$, where only two residues contribute to the ionization rate.
By the way, we denote the quanity $n_0$ as the minimal number of photons required for ionization in circularly polarized laser fields, i.e.
\begin{eqnarray}
\label{eq:n0}
n_0&=&\frac{A_0^2}{2\omega}+\frac{I_p}{\omega}=\frac{2U_p+I_p}{\omega},
\end{eqnarray}
where $U_p=A_0^2/4$ is the pondermotive potential.
Comparing to the case for linearly polarized laser fields, the mean kinetic energy of the electron in a circularly polarized laser field $A_0^2/2=2U_p$ is twice as much.
Applying the residue method for the $k_+$-integral and evaluating the $k_-$-integral, we get the expression for the radial flux at the infinity (see Appendix 3)
\begin{eqnarray}
\label{eq:jfluxlim1}
\lim_{\rho\rightarrow\infty}J_\pm(\rho,t)
&=&\pi \sum_{n\geq n_0}^\infty\int_{-\infty}^\infty dk_y\int_{-\infty}^\infty dk_z\,\frac{h_\pm\left(2\sqrt{k_n^2-k_y^2-k_z^2}\right)-h_\pm\left(-2\sqrt{k_n^2-k_y^2-k_z^2}\right)}{k_n^2-k_y^2-k_z^2},\qquad
\end{eqnarray}
where
\begin{eqnarray}
\label{eq:kn}
\frac{k_n^2}{2}&=&(n-n_0)\omega.
\end{eqnarray}
Using Eq.\,(\ref{eq:hk}) for $k_-=0$ and $n=n_1=n_2$, i.e.
\begin{eqnarray}
h_\pm(k_+)&=&\left(k_+-2A_0\sin(\omega t)\right)\left|F_{n\pm}\left(\left(\frac{k_+}{2},k_y,k_z\right),\omega\right)\right|^2,
\end{eqnarray}
time-averaging over a laser cycle, and using the relations for the $\delta$ function $\delta(\alpha^2-x^2)=[\delta(\alpha-x)+\delta(\alpha+x)]/(2|\alpha|)$ and $\delta(\alpha x)=\delta(x)/|\alpha|$,
we obtain the final formula for the ionization rate from Eq.\,(\ref{eq:ionrate}) as a sum over multiphoton channels
\begin{eqnarray}
\label{eq:w}
w_\pm(\mathcal{E},\omega)
&=&\sum_{n\geq n_0}^\infty w_{n\pm}(\mathcal{E},\omega),
\end{eqnarray}
and
\begin{eqnarray}
\label{eq:wn}
w_{n\pm}(\mathcal{E},\omega)
&=&2\pi \int d\mathbf{k}\,\delta\left(\frac{k^2}{2}-\frac{k_n^2}{2}\right)\left|F_{n\pm}\left(\mathbf{k},\omega\right)\right|^2,
\end{eqnarray}
which coincide exactly with Eqs.\,(13) and (14) of Ref.\,\cite{PPT2} for circular polarization, respectively.
In Eq.\,(\ref{eq:wn}), we recognize that there is the conservation law, namely $k=k_n$, that means that the electron kinetic energy is equal the photon energy minus
the mean electron energy in a circularly polarized laser field and the ionization energy, cf.\,Eqs.\,(\ref{eq:n0}) and (\ref{eq:kn}), i.e.
\begin{eqnarray}
\label{eq:conservation}
\frac{k^2}{2}&=&\frac{k_n^2}{2}=n\omega-2U_p-I_p.
\end{eqnarray}

\subsection{Gauge invariance}

The length gauge was used in the derivation as in the original PPT approach.
However, we note that the result for ionization rate is independent of the gauge.
Rewriting Eqs.\,(\ref{eq:TDSE})--(\ref{eq:green}) using the velocity gauge yields substitution of the original wavefunction $\psi_\pm(\mathbf{r},t)$ in the length gauge by $\psi_\pm(\mathbf{r},t)e^{-i\mathbf{A}_\pm(t)\mathbf{r}}$ in Eq.\,(\ref{eq:psi1}). Thus, in right hand side of Eq.\,(\ref{eq:psi1})  the term $e^{i\mathbf{v}_\pm(t)\mathbf{r}}$ is then replaced by $e^{i\mathbf{v}_\pm(t)\mathbf{r}}e^{-i\mathbf{A}_\pm(t)\mathbf{r}}=e^{i\mathbf{k}\mathbf{r}}$ and the function
in Eq.\,(\ref{eq:f}) is therefore time-independent, i.e.\,$f_\pm(k_\rho,\theta,\phi)=k_\rho\cos(\theta-\phi)$.
Following the derivation in the previous section, Eqs.\,(\ref{eq:appendix1}) and (\ref{eq:hk}) do not have any time-dependent terms anymore, yielding the time-independent radial flux in Eq.\,(\ref{eq:jfluxlim1}).
Therefore, time-averaging over a laser cycle is unnecessary in the velocity gauge, yielding the same result for the ionization rate as in Eqs.\,(\ref{eq:w}) and (\ref{eq:wn}).

\subsection{Derivation of the formula for the probability of the $n$-photon process}

The function $\left|F_{n\pm}\left(\mathbf{k},\omega\right)\right|^2$ in Eq.\,(\ref{eq:wn}) describes the probability of the $n$-photon process in circularly polarized fields,
which is derived in this work not only for $s$- but also for all atomic orbitals, thus beyond the derivations in Refs.\,\cite{PPT1,PPT2}.
With Eqs.\,(\ref{eq:Fkt}) and (\ref{eq:Fouriercoeff}), we start with the general formula for the probability of the $n$-photon process at $k=k_n$
\begin{eqnarray}
\label{eq:Fgeneral}
\left|F_{n\pm}(\mathbf{k},\omega)\right|^2_{k=k_n}&=&\frac{\omega^2}{4\pi^2}\left|\int_{-\pi/\omega}^{\pi/\omega} dt\,\phi_{lm}(\mathbf{v}_\pm(t))e^{iS_\pm(\mathbf{k},t)}\right|^2_{k=k_n},
\end{eqnarray}
where the action $S_{\pm}(\mathbf{k},t)$ at $k=k_n$ is
\begin{eqnarray}
\label{eq:S}
\left.S_{\pm}(\mathbf{k},t)\right|_{k=k_n}&=&\left.\mathbf{k}\left(\boldsymbol\xi_\pm(t)-\boldsymbol\xi_\pm(0)\right)\right|_{k=k_n}+n\omega t.
\end{eqnarray}
It is also obtained by the well-known expression for the action
\begin{eqnarray}
S_{\pm}(\mathbf{k},t)&=&\frac12\int_0^t\mathbf{v}_\pm(\tau)^2\,d\tau+I_pt,
\end{eqnarray}
the conservation law (\ref{eq:conservation}), and the relation
\begin{eqnarray}
\frac12\int_0^t\mathbf{v}_\pm(\tau)^2\,d\tau
&=&\left(\frac{k^2}{2}+2U_p\right)t+\mathbf{k}(\boldsymbol\xi_\pm(t)-\boldsymbol\xi_\pm(0)),
\end{eqnarray}
cf.\,Eq.\,(\ref{eq:intp2}).
Using the saddle point method applied for $\omega\ll I_p$ and $\omega\ll U_p$, we obtain the simple expression for the probability of the $n$-photon process
\begin{eqnarray}
\label{eq:Fgeneral2}
\left|F_{n\pm}(\mathbf{k},\omega)\right|^2_{k=k_n}&=&\frac{\omega^2}{4\pi^2}\left|\phi_{lm}(\mathbf{v}_\pm(t_i))e^{iS_{\pm}(\mathbf{k},t_i)}\sqrt{\frac{2\pi}{S_{\pm}''(\mathbf{k},t_i)}}\right|^2_{k=k_n}.
\end{eqnarray}
It means that the integral in Eq.\,(\ref{eq:Fgeneral}) is accumulated mostly in the small region around the
so-called (complex) ionization time $t_i$ which is uniquely linked to $\mathbf{k}$ and determined by the saddle point equation
\begin{eqnarray}
\label{eq:saddle}
\left.\frac{\partial}{\partial t}\,S_{\pm}(\mathbf{k},t)\right|_{k=k_n,t=t_i}&=&\left.\frac{\mathbf{v}_\pm(t_i)^2}{2}\right|_{k=k_n}+I_p=\left.\frac{\partial}{\partial t}\,\mathbf{k}\boldsymbol\xi_\pm(t)\right|_{k=k_n,t=t_i}+n\omega=0.
\end{eqnarray}
With
\begin{eqnarray}
\label{eq:kxi}
\left.\mathbf{k}\boldsymbol\xi_{\pm}(t)\right|_{k=k_n}&=&\frac{A_0\sqrt{k_n^2-k_z^2}}{\omega}\,\cos(\omega t\mp\theta),
\end{eqnarray}
the saddle point equation (\ref{eq:saddle}) is rewritten as
\begin{eqnarray}
\label{eq:saddle2}
\sin(\omega t_i\mp\theta)&=&\chi_n(k_z),
\end{eqnarray}
where
\begin{eqnarray}
\label{eq:chinkz}
\chi_n(k_z)&=&\frac{n\omega}{A_0\sqrt{k_n^2-k_z^2}}\geq\frac{n\omega}{A_0k_n}=\chi_n(k_z=0)=:\chi_n.
\end{eqnarray}
Using Eqs.\,(\ref{eq:n0}) and (\ref{eq:kn}), the variable $\chi_n$ is reexpressed as
\begin{eqnarray}
\label{eq:chin0}
\chi_n&=&\sqrt{\frac{n^2\omega}{2A_0^2(n-n_0)}}=\sqrt{\frac{n^2(1+\gamma^2)}{4n_0(n-n_0)}},
\end{eqnarray}
where $\gamma=\sqrt{2I_p}/A_0>0$ is the Keldysh parameter \cite{keldysh}
discriminating between adiabatic tunneling ($\gamma\ll 1$), non-adiabatic tunneling ($\gamma\sim 1$) \cite{YudinIvanov}, and multiphoton ionization ($\gamma\gg 1$).
Eq.\,(\ref{eq:chin0}) is further rewritten as
\begin{eqnarray}
\label{eq:chin0t}
\chi_n&=&\sqrt{\frac{1+\gamma^2}{1-\zeta^2}},
\end{eqnarray}
where
\begin{eqnarray}
\label{eq:zeta}
\zeta&=&\frac{2n_0}{n}-1\in(-1,1],
\end{eqnarray}
corresponding to the range $n\geq n_0$. In this range, $\chi_n$ is always larger than 1, thus $\chi_n(k_z)>1$, cf.\,Eq.\,(\ref{eq:chinkz}).
Therefore, the ionization time $t_i$ in Eq.\,(\ref{eq:saddle2}) must be complex, i.e.\ $t_i=\mathrm{Re}\,t_i+i\,\mathrm{Im}\,t_i$, and we get two equations for $\mathrm{Re}\,t_i$ and $\mathrm{Im}\,t_i$,
\begin{eqnarray}
\label{eq:saddle_1}
\sin(\omega\,\mathrm{Re}\,t_i\mp\theta)\cosh(\omega\,\mathrm{Im}\,t_i)&=&\chi_n(k_z)\\
\label{eq:saddle_2}
\cos(\omega\,\mathrm{Re}\,t_i\mp\theta)\sinh(\omega\,\mathrm{Im}\,t_i)&=&0.
\end{eqnarray}
The corresponding solutions for $\mathrm{Im}\,t_i\neq 0$ are
\begin{eqnarray}
\label{eq:solution1}
\omega\,\mathrm{Re}\,t_i&=&\frac{\pi}{2}\pm\theta+2\pi N,\\
\label{eq:solution2}
\omega\,\mathrm{Im}\,t_i&=&\mathrm{arcosh}\,\chi_n(k_z),
\end{eqnarray}
where $N\in\mathbb{Z}$ is chosen such that $\omega\,\mathrm{Re}\,t_i$ lies in the interval of a laser cylce, that is between $-\pi$ and $\pi$, cf.\,Eq.\,(\ref{eq:Fgeneral}).
The complex time $t_i$ can be interpreted as the time of entering into the barrier,
while its imaginary $\mathrm{Im}\,t_i$ and real $\mathrm{Re}\,t_i$ parts are the tunneling time and the time of exiting the barrier, respectively \cite{PPT3}.
With Eqs.\,(\ref{eq:kxi}), (\ref{eq:chinkz}), (\ref{eq:solution1}), (\ref{eq:solution2}), and $\sinh x=\sqrt{\cosh^2 x-1}$, the corresponding action $S_{\pm}(\mathbf{k},t_i)$ at $k=k_n$ (Eq.\,\ref{eq:S}) is
\begin{eqnarray}
\label{eq:Sti}
\left.S_{\pm}(\mathbf{k},t_i)\right|_{k=k_n}&=&n\left(\frac{\pi}{2}\pm \theta+2\pi N-\frac{\cos\theta}{\chi_n(k_z)}\right)+in\left(\mathrm{arcosh}\,\chi_n(k_z)-\sqrt{1-\frac{1}{\chi_n(k_z)^2}}\right).\qquad
\end{eqnarray}
Only its imaginary part
\begin{eqnarray}
\label{eq:Stiim}
\left.\mathrm{Im}\,S(\mathbf{k},t_i)\right|_{k=k_n}&=&n\left(\mathrm{arcosh}\,\chi_n(k_z)-\sqrt{1-\frac{1}{\chi_n(k_z)^2}}\right)
\end{eqnarray}
does not depend on the sense $(\pm)$ of circular polarization, thus we can omit the index $\pm$ of the imaginary part of the action.
It means that the imaginary actions are equal for right and left circular polarizations.
Using Eqs.\,(\ref{eq:S}), (\ref{eq:kxi})--(\ref{eq:chinkz}), the absolute value of the second derivative of the action $S_{\pm}''(\mathbf{k},t_i)$ in Eq.\,(\ref{eq:Fgeneral2}) at $k=k_n$
\begin{eqnarray}
\label{eq:Ssecond}
\left|\frac{\partial^2}{\partial t^2}\,S(\mathbf{k},t)\right|_{k=k_n,t=t_i}&=&n\omega^2\,\sqrt{1-\frac{1}{\chi_n(k_z)^2}}
\end{eqnarray}
is also independent of the sense $(\pm)$ of circular polarization.
Then, the expression for the probability of the $n$-photon process (\ref{eq:Fgeneral2}) is rewritten as
\begin{eqnarray}
\label{eq:Fgeneral3}
\left|F_{n\pm}(\mathbf{k},\omega)\right|^2_{k=k_n}&=&\frac{\left|\phi_{lm}(\mathbf{v}_\pm(t_i))\right|^2_{k=k_n}}{2\pi n\sqrt{1-1/\chi_n(k_z)^2}}\,e^{-2n\left(\mathrm{arcosh}\,\chi_n(k_z)-\sqrt{1-1/\chi_n(k_z)^2}\right)}.
\end{eqnarray}
Thus, the dependence of the ionization rate on the sense of circular polarization is only due to the prefactor
\begin{eqnarray}
\label{eq:philmti}
\left|\phi_{lm}(\mathbf{v}_\pm(t_i))\right|^2_{k=k_n}&=&\frac14\left|(\mathbf{v}_\pm(t_i)^2+2I_p)\tilde\varphi_{lm}(\mathbf{v}_\pm(t_i))\right|^2_{k=k_n},
\end{eqnarray}
cf.\ Eq.\,(\ref{eq:philm}).
Using Eqs.\,(\ref{eq:p}), (\ref{eq:solution1}), and (\ref{eq:solution2}), the initial electron velocity
\begin{eqnarray}
\label{eq:pti}
\left.\mathbf{v}_\pm(t_i)\right|_{k=k_n}
&=&\left.v_{x\pm}(t_i)\right|_{k=k_n}\,\mathbf{e}_x+\left.v_{y\pm}(t_i)\right|_{k=k_n}\,\mathbf{e}_y+k_z\,\mathbf{e}_z
\end{eqnarray}
with $x$- and $y$-components
\begin{eqnarray}
\label{eq:pxti}
\left.v_{x\pm}(t_i)\right|_{k=k_n}&=&k_\rho\cos\theta-A_0(\chi_n(k_z)\cos\theta\mp i\sqrt{\chi_n(k_z)^2-1}\,\sin\theta)\\
\label{eq:pyti}
\left.v_{y\pm}(t_i)\right|_{k=k_n}&=&k_\rho\sin\theta-A_0(\chi_n(k_z)\sin\theta\pm i\sqrt{\chi_n(k_z)^2-1}\,\cos\theta)
\end{eqnarray}
specifices the required momentum of the initial wavefunction, i.e.\ $\tilde\varphi_{lm}(\mathbf{v}_\pm(t_i))$.
The amount of this momentum depends on the orbital, in particular it is different for $p_+$ and $p_-$ orbitals.
Because of the saddle point equation (\ref{eq:saddle}), i.e.\ $\mathbf{v}_\pm(t_i)^2|_{k=k_n}+2I_p=0$, the initial wavefunction $\tilde\varphi_{lm}(\mathbf{v}_\pm(t_i))$ must have the pole at
$\mathbf{v}_\pm(t_i)^2|_{k=k_n}=-2I_p=-\kappa^2$ that yields non-zero prefactor $\left|\phi_{lm}(\mathbf{v}_\pm(t_i))\right|^2_{k=k_n}$, cf.\ Eq.\ (\ref{eq:philmti}).
For short-range potentials, it corresponds to the wavefunction in coordinate representation asymptotically far from the core \cite{PPT1}, i.e.
\begin{eqnarray}
\label{eq:varphicoord}
\varphi_{lm}(\mathbf{r})&=& C_{\kappa l}\kappa^{3/2}\,\frac{e^{-\kappa r}}{\kappa r}\,Y_{lm}(\theta_r,\phi_r),
\end{eqnarray}
with the constant $C_{\kappa l}$, depending on $\kappa=\sqrt{2I_p}$ and $l$ as well as details of the potential near the core.
Using spherical harmonics
\begin{eqnarray}
\label{eq:harmonics}
Y_{lm}(\theta_r,\phi_r)&=&\sqrt{\frac{2l+1}{4\pi}\frac{(l-m)!}{(l+m)!}}\,P_l^m(\cos\theta_r)e^{im\phi_r},
\end{eqnarray}
Fourier transformation (\ref{eq:phip}), and
\begin{eqnarray}
\label{eq:vt}
-i\mathbf{v}_\pm(t_i)\mathbf{r}=-iv_\pm(t_i)r(\sin\theta_{v\pm}(t_i)\sin\theta_r\cos(\phi_{v\pm}(t_i)-\phi_r)+\cos\theta_{v\pm}(t_i)\cos\theta_r)
\end{eqnarray}
in spherical coordinates, we evaluate two integrals over $\phi_r$ and $\theta_r$ with the help of the Bessel function and Ref.\ \cite{neves} to yield
the intermediate result (see Appendix 4)
\begin{eqnarray}
\label{eq:appendix4a}
\tilde\varphi_{lm}(\mathbf{v}_\pm(t_i))
&=&C_{\kappa l}\sqrt{\frac{\kappa}{v_\pm(t_i)}}\,Y_{lm}(\theta_{v\pm}(t_i),\phi_{v\pm}(t_i))\,e^{-il\pi/2}\int_0^\infty dr\,\sqrt{r}\,e^{-\kappa r}J_{l+1/2}(v_\pm(t_i)r).\qquad
\end{eqnarray}
Expanding the Bessel function in Taylor series and using the Gamma function $\Gamma(z)$, the integration over $r$ is easily carried out.
The resulting series is then compacted as the Gaussian hypergeometric series (see Appendix 4)
\begin{eqnarray}
\label{eq:varphiseries}
\tilde\varphi_{lm}(\mathbf{v}_\pm(t_i))
&=&\frac{C_{\kappa l}}{\sqrt{2\kappa^3}}\left(\frac{v_\pm(t_i)}{2\kappa}\right)^lY_{lm}(\theta_{v\pm}(t_i),\phi_{v\pm}(t_i))\,e^{-il\pi/2}\\\nonumber
&&\frac{\Gamma(l+2)}{\Gamma(l+3/2)}\,_2F_1\left(\frac{l}{2}+1,\frac{l}{2}+\frac32;l+\frac32;-\frac{v_\pm(t_i)^2}{\kappa^2}\right).
\end{eqnarray}
The hypergeometric series does not converge at the saddle point $v_\pm(t_i)^2|_{k=k_n}=-\kappa^2$, thus this series has the pole as expected above.
Multiplying Eq.\,(\ref{eq:varphiseries}) by $v_\pm(t_i)^2+\kappa^2$ yields the series (see Appendix 4)
\begin{eqnarray}
\label{eq:varphifinal}
(v_\pm(t_i)^2+\kappa^2)\tilde\varphi_{lm}(\mathbf{v}_\pm(t_i))
&=&C_{\kappa l}\,\sqrt{\frac{2\kappa}{\pi}}\left(\frac{v_\pm(t_i)}{\kappa}\right)^lY_{lm}(\theta_{v\pm}(t_i),\phi_{v\pm}(t_i))\,e^{-il\pi/2}\\\nonumber
&&\frac{\sqrt{\pi}}{2^{l+1}}\,\frac{\Gamma(l+2)}{\Gamma(l+3/2)}\,_2F_1\left(\frac{l}{2}+\frac12,\frac{l}{2};l+\frac32;-\frac{v_\pm(t_i)^2}{\kappa^2}\right),
\end{eqnarray}
which is convergent at the saddle point, i.e.
\begin{eqnarray}
\label{eq:2F1convergent}
\frac{\sqrt{\pi}}{2^{l+1}}\,\frac{\Gamma(l+2)}{\Gamma(l+3/2)}\,_2F_1\left(\frac{l}{2},\frac{l}{2}+\frac12;l+\frac32;1\right)&=&1.
\end{eqnarray}
Thus, the prefactor (\ref{eq:philmti}) is simplified as
\begin{eqnarray}
\label{eq:prefactor}
\left|\phi_{lm}(\mathbf{v}_\pm(t_i))\right|^2_{k=k_n}&=&\frac{|C_{\kappa l}|^2\sqrt{2I_p}}{2\pi}\left|Y_{lm}(\theta_{v\pm}(t_i),\phi_{v\pm}(t_i))\right|^2_{k=k_n},
\end{eqnarray}
cf.\ Ref.\ \cite{PPT1}. With Eqs.\,(\ref{eq:Fgeneral3}) and (\ref{eq:harmonics}), we finally obtain the general result for the probability of the $n$-photon process for all atomic orbitals
\begin{eqnarray}
\label{eq:Fgeneral4}
\left|F_{n\pm}(\mathbf{k},\omega)\right|^2_{k=k_n}&=&\frac{|C_{\kappa l}|^2\sqrt{2I_p}\,(2l+1)}{16\pi^3 n\sqrt{1-1/\chi_n(k_z)^2}}\frac{(l-|m|)!}{(l+|m|)!}
\left|\,P_l^{|m|}\left(\frac{ik_z}{\sqrt{2I_p}}\right)\right|^2\left|e^{im\phi_{v\pm}(t_i)}\right|^2_{k=k_n}\\\nonumber
&&e^{-2n\left(\mathrm{arcosh}\,\chi_n(k_z)-\sqrt{1-1/\chi_n(k_z)^2}\right)},
\end{eqnarray}
where $\cos\theta_{v\pm}(t_i)|_{k=k_n}=v_z/v_\pm(t_i)|_{k=k_n}=\pm ik_z/\sqrt{2I_p}$ was used.
The square of the associated Legendre polynomials in Eq.\,(\ref{eq:Fgeneral4}) are equal to $1$ for $s$ orbitals, $k_z^2/(2I_p)$ for $p_0$ orbitals, $(k_z^2+2I_p)/(2I_p)$ for $p_\pm$ orbitals, and so on.
The ionizaztion rates for orbitals with $m=0$ (e.g.\ $s$ and $p_0$ orbitals) are independent of the sense of circular polarization.
For $m\neq 0$ (e.g.\ $p_\pm$ orbitals), they depend on the polarization sense, solely due to the factor $|e^{im\phi_{v\pm}(t_i)}|_{k=k_n}^2$.
For $m\neq 0$, this factor is not equal to unity, because the so-called tunneling momentum angle $\phi_{v\pm}(t_i)$, which is related by
\begin{eqnarray}
\label{eq:cosv}
\cos\phi_{v\pm}(t_i)&=&\frac{v_{x\pm}(t_i)}{v_{\rho\pm}(t_i)}\\
\label{eq:sinv}
\sin\phi_{v\pm}(t_i)&=&\frac{v_{y\pm}(t_i)}{v_{\rho\pm}(t_i)},
\end{eqnarray}
is complex.
With $v_{\rho\pm}(t_i)^2|_{k=k_n}=-(k_z^2+2I_p)$, $\gamma=\sqrt{2I_p}/A_0$, Eqs.\,(\ref{eq:n0}), (\ref{eq:chinkz}), (\ref{eq:pxti}), and (\ref{eq:pyti}),
the factor $|e^{im\phi_{v\pm}(t_i)}|_{k=k_n}^2$ for $m=\pm 1$ is (see Appendix 5)
\begin{eqnarray}
\left|e^{im\phi_{v\pm}(t_i)}\right|_{k=k_n}^2
&=&\left|\cos\phi_{v\pm}(t_i)+i\,\mathrm{sgn}(m)\sin\phi_{v\pm}(t_i)\right|_{k=k_n}^2\\
\label{eq:expim}
&=&\frac{I_p\left[2\chi_n(k_z)^2\left(1\mp\,\mathrm{sgn}(m)\sqrt{1-1/\chi_n(k_z)^2}\right)-(1+\gamma^2)n/n_0\right]^2}{2\gamma^2\chi_n(k_z)^2\left(k_z^2+2I_p\right)}.
\end{eqnarray}
Now, we can see that the ionization rates for a given circular polarization are different for orbitals with opposite quantum numbers $m=\pm 1$ (e.g.\ $p_\pm$ orbitals).
Because of the term $\mp\mathrm{sgn}(m)$ in Eq.\,(\ref{eq:expim}), the ionization rate for the $p_+$ (or $p_-$) orbital and right ciruclar polarization is
the same as the ionization rate for the $p_-$ (or $p_+$) orbital and left circular polarization, supporting the fundamental symmetry in electrodynamics.

\subsection{Accurate formulas for the time-averaged ionization rates for $s$ and $p$ orbitals}

Since the function $\left|F_{n\pm}(\mathbf{k},\omega)\right|^2_{k=k_n}$ (Eq.\,(\ref{eq:Fgeneral4})) depends only on $k_z^2$,
the two integrations over $k_\rho$ and $\theta$ in the formula for the time-averaged ionization rate (Eqs.\,(\ref{eq:w}) and (\ref{eq:wn})) are easily carried out and the result is simplified to
\begin{eqnarray}
\label{eq:wz}
w_\pm(\mathcal{E},\omega)
&=&8\pi^2\sum_{n\geq n_0}^\infty \int_0^{k_n} dk_z\,
\left|F_{n\pm}\left(\mathbf{k},\omega\right)\right|^2_{k=k_n}.
\end{eqnarray}
With Eqs.\,(\ref{eq:Fgeneral4}) and (\ref{eq:expim}), the accurate formulas for the time-averaged ionization rates are
\begin{eqnarray}
\label{eq:wsacc}
w^s(\mathcal{E},\omega)
&=&\frac{|C_{\kappa 0}|^2\sqrt{2I_p}}{2\pi}\sum_{n\geq n_0}^\infty \frac{1}{n}\int_0^{k_n} dk_z\,
\frac{e^{-2n\left(\mathrm{arcosh}\,\chi_n(k_z)-\sqrt{1-1/\chi_n(k_z)^2}\right)}}{\sqrt{1-1/\chi_n(k_z)^2}}
\end{eqnarray}
for $s$ orbitals,
\begin{eqnarray}
\label{eq:wp0acc}
w^{p_0}(\mathcal{E},\omega)
&=&\frac{3|C_{\kappa 1}|^2}{2\pi\sqrt{2I_p}}\sum_{n\geq n_0}^\infty \frac{1}{n}\int_0^{k_n} dk_z\,
\frac{k_z^2\,e^{-2n\left(\mathrm{arcosh}\,\chi_n(k_z)-\sqrt{1-1/\chi_n(k_z)^2}\right)}}{\sqrt{1-1/\chi_n(k_z)^2}}
\end{eqnarray}
for $p_0$ orbitals, and
\begin{eqnarray}
\label{eq:wp1acc}
w_\pm^{p_\pm}(\mathcal{E},\omega)
&=&\frac{3|C_{\kappa 1}|^2\sqrt{2I_p}}{16\pi\gamma^2}\sum_{n\geq n_0}^\infty\frac{1}{n} \int_0^{k_n} dk_z\,
\frac{e^{-2n\left(\mathrm{arcosh}\,\chi_n(k_z)-\sqrt{1-1/\chi_n(k_z)^2}\right)}}{\chi_n(k_z)^2\sqrt{1-1/\chi_n(k_z)^2}}\\\nonumber
&&\qquad\qquad\left[2\chi_n(k_z)^2\left(1\mp\,\mathrm{sgn}(m)\sqrt{1-1/\chi_n(k_z)^2}\right)-(1+\gamma^2)n/n_0\right]^2
\end{eqnarray}
for $p_\pm$ orbitals.

\subsection{Approximate formulas for the time-averaged ionization rates for $s$ and $p$ orbitals}

In Eqs.\,(\ref{eq:wsacc})--(\ref{eq:wp1acc}), the exponential function has the maximum at $k_z=0$, confirming our expectation that the electron leaves mostly in the polarization plane of the laser field,
i.e.\ $x/y$-plane, and that the electron ionization along the propagation axis ($z$-axis) is suppressed.
Therefore, we use Taylor series of the exponent at $k_z\approx 0$ up to second order, i.e.
\begin{eqnarray}
\label{eq:taylorexp}
\mathrm{arcosh}\,\chi_n(k_z)-\sqrt{1-1/\chi_n(k_z)^2}&\approx&
\mathrm{arcosh}\,\chi_n-\sqrt{1-1/\chi_n^2}+\frac12\sqrt{1-1/\chi_n^2}\left(\frac{k_z}{k_n}\right)^2.\qquad
\end{eqnarray}
The $k_z$-dependent prefactors in Eqs.\,(\ref{eq:wsacc})--(\ref{eq:wp1acc}) are then replaced by the non-vanishing lowest-order terms of the corresponding Taylor series at $k_z\approx0$, i.e.
\begin{eqnarray}
\label{eq:kzpres}
\frac{1}{\sqrt{1-1/\chi_n(k_z)^2}}&\approx&\frac{1}{\sqrt{1-1/\chi_n^2}}
\end{eqnarray}
for $s$ orbitals,
\begin{eqnarray}
\label{eq:kzprep0}
\frac{k_z^2}{\sqrt{1-1/\chi_n(k_z)^2}}&\approx&\frac{k_z^2}{\sqrt{1-1/\chi_n^2}}
\end{eqnarray}
for $p_0$ orbitals, and
\begin{eqnarray}
\label{eq:kzprep1}
&&\frac{\left[2\chi_n(k_z)^2\left(1\mp\,\mathrm{sgn}(m)\sqrt{1-1/\chi_n(k_z)^2}\right)-(1+\gamma^2)n/n_0\right]^2}{\chi_n(k_z)^2\sqrt{1-1/\chi_n(k_z)^2}}\\\nonumber
&\approx&\frac{\left[2\chi_n^2\left(1\mp\,\mathrm{sgn}(m)\sqrt{1-1/\chi_n^2}\right)-(1+\gamma^2)n/n_0\right]^2}{\chi_n^2\sqrt{1-1/\chi_n^2}}
\end{eqnarray}
for $p_\pm$ orbitals.
For $p_0$ orbitals, the $k_z$-dependent prefactor in Eq.\,(\ref{eq:kzprep0}) has no zeroth-order term due to the existence of the factor $k_z^2$.
That means that the ionization rate for $p_0$ orbitals in the polarization plane is zero because of the destructive interference coming from two phase-opposite lobes.
It also concludes that the approximation $k_z\approx 0$ for $p_0$ orbitals may be not very appropriate, i.e.\ the electron from the $p_0$ orbital will leave with non-zero final momentum component $k_z\neq 0$,
i.e.\ parallel to the $z$-axis, due to its orbital shape.
With the Taylor approximations (\ref{eq:taylorexp})--(\ref{eq:kzprep1}), we evaluate the remaining integrals in Eqs.\,(\ref{eq:wsacc})--(\ref{eq:wp1acc}) over $k_z$ as
\begin{eqnarray}
\label{eq:int1}
\int_0^{k_n} dk_z\,e^{-a_n^2\left(\frac{k_z}{k_n}\right)^2}
&=&\frac{\sqrt{\pi}\,k_n\,\mathrm{erf}\left(a_n\right)}{2a_n}
\end{eqnarray}
and
\begin{eqnarray}
\label{eq:int2}
\int_0^{k_n} dk_z\,k_z^2\,e^{-a_n^2\left(\frac{k_z}{k_n}\right)^2}
&=&k_n^3\left(\frac{\sqrt{\pi}\,\mathrm{erf}\left(a_n\right)}{4 a_n^3}-\frac{e^{-a_n^2}}{2 a_n^2}\right)
\end{eqnarray}
where $a_n=\sqrt{n}\,(1-1/\chi_n^2)^{1/4}$.
For $\omega\ll I_p$, the minimal number of photons $n_0$ (Eq.\,(\ref{eq:n0})) must be very large.
It follows that $n\geq n_0\gg 1$, hence $\chi_n\gg 1$ (cf.\ Eqs.\,(\ref{eq:chin0t}) and (\ref{eq:zeta})) and $a_n\gg 1$.
In the limit $a_n\rightarrow\infty$, the error function $\mathrm{erf}(a_n)$ and the exponential function $e^{-a_n^2}$ tend to unity and zero, respectively.
For $\omega\ll I_p$, we use Eqs.\,(\ref{eq:wsacc})--(\ref{eq:int2}) and Eqs.\,(\ref{eq:n0}), (\ref{eq:chinkz}), $A_0=\sqrt{2I_p}/\gamma$,
i.e.\ $k_n=n\omega\gamma/(\chi_n\sqrt{2I_p})$ and $\omega=I_p(1+\gamma^2)/(n_0\gamma^2)$,
to obtain the approximate formuals for the time-averaged ionization rates
\begin{eqnarray}
\label{eq:wsapp}
w^s(\mathcal{E},\omega)
&=&\frac{|C_{\kappa 0}|^2I_p(1+\gamma^2)}{4\sqrt{\pi}\,n_0\gamma}\sum_{n\geq n_0}^\infty
\frac{e^{-2n\left(\mathrm{arcosh}\,\chi_n-\sqrt{1-1/\chi_n^2}\right)}}{\sqrt{n}\,\chi_n\left(1-1/\chi_n^2\right)^{3/4}}
\end{eqnarray}
for $s$ orbitals,
\begin{eqnarray}
\label{eq:wp0app}
w^{p_0}(\mathcal{E},\omega)
&=&\frac{3|C_{\kappa 1}|^2I_p(1+\gamma^2)^3}{32\sqrt{\pi}\,n_0^3\gamma^3}\sum_{n\geq n_0}^\infty
\frac{\sqrt{n}\,e^{-2n\left(\mathrm{arcosh}\,\chi_n-\sqrt{1-1/\chi_n^2}\right)}}{\chi_n^3\left(1-1/\chi_n^2\right)^{5/4}}
\end{eqnarray}
for $p_0$ orbitals, and
\begin{eqnarray}
\label{eq:wp1app}
w_\pm^{p_\pm}(\mathcal{E},\omega)
&=&\frac{3|C_{\kappa 1}|^2I_p(1+\gamma^2)}{8\sqrt{\pi}\,n_0\gamma^3}\sum_{n\geq n_0}^\infty\frac{\chi_n\,e^{-2n\left(\mathrm{arcosh}\,\chi_n-\sqrt{1-1/\chi_n^2}\right)}}{\sqrt{n}\,(1-1/\chi_n^2)^{3/4}}\\\nonumber
&&\qquad\qquad\qquad\qquad\qquad\left[\sqrt{1-1/\chi_n^2}\mp(2n_0/n-1)\,\mathrm{sgn}(m)\right]^2
\end{eqnarray}
for $p_\pm$ orbitals.
These expressions Eqs.\,(\ref{eq:wsapp})--(\ref{eq:wp1app}) depend on $\chi_n$ and $n$.
If we use Eqs.\,(\ref{eq:chin0t}) and (\ref{eq:zeta}) as well as $\mathrm{arcosh}\,\chi_n=\mathrm{artanh}\sqrt{1-1/\chi_n^2}$, then we obtain alternative expressions for the ionization rates depending on $\zeta$, i.e.
\begin{eqnarray}
\label{eq:wsappalt}
w^s(\mathcal{E},\omega)
&=&\frac{|C_{\kappa 0}|^2 I_p}{4\sqrt{2\pi}\,n_0^{3/2}}\left(1+\frac{1}{\gamma^2}\right)^{1/2}\\\nonumber
&&\sum_{n\geq n_0}^\infty\left(1+\zeta\right)\sqrt{1-\zeta}\left(\frac{1+\gamma^2}{\zeta^2+\gamma^2}\right)^{3/4}
e^{-\frac{4n_0}{1+\zeta}\left(\mathrm{artanh}\,\sqrt{\frac{\zeta^2+\gamma^2}{1+\gamma^2}}-\sqrt{\frac{\zeta^2+\gamma^2}{1+\gamma^2}}\right)}
\end{eqnarray}
for $s$ orbitals, where it coincides exactly with Eqs.\,(68) and (69) of Ref.\,\cite{PPT1},
\begin{eqnarray}
\label{eq:wp0appalt}
w^{p_0}(\mathcal{E},\omega)
&=&\frac{3|C_{\kappa 1}|^2I_p}{16\sqrt{2\pi}\,n_0^{5/2}}\left(1+\frac{1}{\gamma^2}\right)^{3/2}\\\nonumber
&&\sum_{n\geq n_0}^\infty
\left(1-\zeta^2\right)
\sqrt{1-\zeta}
\left(\frac{1+\gamma^2}{\zeta^2+\gamma^2}\right)^{5/4}
e^{-\frac{4n_0}{1+\zeta}\left(\mathrm{artanh}\,\sqrt{\frac{\zeta^2+\gamma^2}{1+\gamma^2}}-\sqrt{\frac{\zeta^2+\gamma^2}{1+\gamma^2}}\right)}
\end{eqnarray}
for $p_0$ orbitals, and
\begin{eqnarray}
\label{eq:wp1appalt}
w_\pm^{p_\pm}(\mathcal{E},\omega)
&=&\frac{3|C_{\kappa 1}|^2I_p}{8\sqrt{2\pi}\,n_0^{3/2}}\left(1+\frac{1}{\gamma^2}\right)^{3/2}
\sum_{n\geq n_0}^\infty
\left(\sqrt{\frac{\zeta^2+\gamma^2}{1+\gamma^2}}\mp\zeta\,\mathrm{sgn}(m)\right)^2\\\nonumber
&&\frac{1}{\sqrt{1-\zeta}}\left(\frac{1+\gamma^2}{\zeta^2+\gamma^2}\right)^{3/4}
e^{-\frac{4n_0}{1+\zeta}\left(\mathrm{artanh}\,\sqrt{\frac{\zeta^2+\gamma^2}{1+\gamma^2}}-\sqrt{\frac{\zeta^2+\gamma^2}{1+\gamma^2}}\right)}\\\nonumber
\end{eqnarray}
for $p_\pm$ orbitals.
Since $\left|\sqrt{(\zeta^2+\gamma^2)/(1+\gamma^2)}\right|\geq|\zeta|$, we recognize in Eq.\,(\ref{eq:wp1appalt}) that the $n$-photon ionization rate is maximal
for $\mp\mathrm{sgn}(m)=1$ if $\zeta>0$ and for $\pm\mathrm{sgn}(m)=1$ if $\zeta<0$.
Therefore, for $\zeta>0$ (low photon and kinetic energies) and e.g.\ for right circular polarization, the rate for $p_-$ orbitals is larger than the one for $p_+$ orbitals.
For $\zeta<0$ (high photon and kinetic energies), however, the rate for $p_+$ orbitals is larger than the one for $p_-$ orbitals.
For $\zeta=0$, corresponding to the photon energy $n\omega=2n_0\omega=4U_p+2I_p$ and electron kinetic energy $k_n^2/2=(n-n_0)\omega=n_0\omega=2U_p+I_p$,
the ionization rates for both $p_\pm$ orbitals are identical.
By the way, we would like to stress that the ionization rate for $p_0$ orbitals (Eq.\,(\ref{eq:wp0appalt})) is very small compared to the rates for $s$ and $p_\pm$ orbitals,
due to the ionization supression in the polarization plane.

\subsection{Simple formulas for the time-averaged ionization rates for $s$ and $p$ orbitals}

To obtain the simple analytical expressions for the ionization rates, the summation over $n$-photon processes in Eqs.\,(\ref{eq:wsappalt})--(\ref{eq:wp1appalt}) can be replaced with integration over $\zeta$, i.e.
\begin{eqnarray}
\sum_{n\geq n_0}^\infty&\approx&\int_{n_0}^\infty dn\,\,\,=\,\,\,2n_0\int_{-1}^1 \frac{d\zeta}{(1+\zeta)^2}.
\end{eqnarray}
For $\omega\ll I_p$, i.e.\ $n_0\gg 1$, the saddle point method for integration over $\zeta$ is then applied, where the exponent in Eqs.\,(\ref{eq:wsappalt})--(\ref{eq:wp1appalt})
\begin{eqnarray}
\label{eq:szeta}
S(\zeta,\gamma)&=&-\frac{4n_0}{1+\zeta}\left(\mathrm{artanh}\sqrt{\frac{\zeta^2+\gamma^2}{1+\gamma^2}}-\sqrt{\frac{\zeta^2+\gamma^2}{1+\gamma^2}}\right)
\end{eqnarray}
has a unique maximum at $\zeta=\zeta_0(\gamma)$. This maximum is determined by the saddle point equation
\begin{eqnarray}
\left.\frac{\partial}{\partial \zeta}\,S(\zeta,\gamma)\right|_{\zeta=\zeta_0}&=&0,
\end{eqnarray}
that yields the transcendental equation for $\zeta_0(\gamma)$
\begin{eqnarray}
\label{eq:eqt0g}
\mathrm{artanh}\sqrt{\frac{\zeta_0^2+\gamma^2}{1+\gamma^2}}&=&\frac{1}{1-\zeta_0}\sqrt{\frac{\zeta_0^2+\gamma^2}{1+\gamma^2}},
\end{eqnarray}
or equivalently
\begin{eqnarray}
\label{eq:eqnmax}
\mathrm{artanh}\sqrt{1-1/\chi_{n_{\max}}^2}&=&\frac12\frac{n_{\max}}{n_{\max}-n_0}\sqrt{1-1/\chi_{n_{\max}}^2},
\end{eqnarray}
where $\chi_{n_{\max}}=\sqrt{(1+\gamma^2)/(1-\zeta_0^2)}$ (cf.\ Eq.\,(\ref{eq:chin0t})) and $n_{\max}=2n_0/(1+\zeta_0)$ is the number of photons for which the $n$-photon ionization rate is maximal,
corresponding to the electron kinetic energy $k^2_{n_{\max}}/2=(n_{\max}-n_0)\omega=(2U_p+I_p)(1-\zeta_0)/(1+\zeta_0)$.
Since the solution of Eq.\,(\ref{eq:eqt0g}) is in the positive range $\zeta_0(\gamma)\in[0,1]$, the maximum of the ($n_{\max}$-photon) ionization rate for right (left) circular polarization
is dominated by the electron ionization from the $p_-$ ($p_+$) orbital.
In the adiabatic limit ($\gamma\ll 1$), the solution is approximated as $\zeta_0(\gamma)\approx\gamma^2/3$ (see Ref.\,\cite{PPT1} and Appendix 6),
corresponding to the electron kinetic energy $k^2_{n_{\max}}/2\approx(2U_p+I_p)(1-2\gamma^2/3)\approx2U_p+I_p/3$,
whereas in the non-adiabatic limit ($\gamma\gg 1$), it yields $\zeta_0(\gamma)\approx 1-1/\ln\gamma$, see Ref.\,\cite{PPT1}.
The exponent (Eq.\,(\ref{eq:szeta})) at the saddle point $\zeta_0(\gamma)$ is
\begin{eqnarray}
S(\zeta_0,\gamma)&=&-\frac{2\mathcal{E}_0}{3\mathcal{E}}\,g(\gamma),
\end{eqnarray}
where $\mathcal{E}_0=(2I_p)^{3/2}$ and
\begin{eqnarray}
\label{eq:final_g}
g(\gamma)&=&\frac{3\zeta_0}{\gamma^2(1-\zeta_0^2)}\sqrt{(1+\gamma^2)(\zeta_0^2/\gamma^2+1)},
\end{eqnarray}
and it does not depend on orbitals.
We also need the second derivative of the exponent $S''(\zeta_0,\gamma)$, i.e.
\begin{eqnarray}
\left.\frac{\partial^2}{\partial \zeta^2}\,S(\zeta,\gamma)\right|_{\zeta=\zeta_0}
&=&-\frac{4n_0(2\zeta_0^2+\zeta_0^2\gamma^2+\gamma^2)}{(1+\zeta_0)^3(1-\zeta_0)^2(\zeta_0^2+\gamma^2)}\sqrt{\frac{\zeta_0^2+\gamma^2}{1+\gamma^2}},
\end{eqnarray}
to apply the saddle point method according to
\begin{eqnarray}
\int_{-1}^1 d\zeta\,f(\zeta) e^{S(\zeta,\gamma)}&=&f(\zeta_0)e^{S(\zeta_0,\gamma)}\sqrt{\frac{2\pi}{-S''(\zeta_0,\gamma)}}.
\end{eqnarray}
Therefore, with these equations and Eqs.\,(\ref{eq:wsappalt})--(\ref{eq:wp1appalt}), we obtain the compact expressions for the time-averaged ionization rates
\begin{eqnarray}
\label{eq:wssimple}
w^s(\mathcal{E},\omega)
&=&|C_{\kappa 0}|^2 I_p\,\frac{\mathcal{E}}{2\mathcal{E}_0}\,h^s(\gamma)\,e^{-\frac{2\mathcal{E}_0}{3\mathcal{E}}\,g(\gamma)}
\end{eqnarray}
for $s$ orbitals,
\begin{eqnarray}
\label{eq:wp0simple}
w^{p_0}(\mathcal{E},\omega)
&=&|C_{\kappa 1}|^2 I_p\,\frac{\mathcal{E}}{2\mathcal{E}_0}\,h^{p_0}(\gamma)\,e^{-\frac{2\mathcal{E}_0}{3\mathcal{E}}\,g(\gamma)}
\end{eqnarray}
for $p_0$ orbitals, and
\begin{eqnarray}
\label{eq:wp1simple}
w^{p_\pm}_\pm(\mathcal{E},\omega)
&=&|C_{\kappa 1}|^2 I_p\,\frac{\mathcal{E}}{2\mathcal{E}_0}\,h^{p_\pm}_\pm(\gamma)\,e^{-\frac{2\mathcal{E}_0}{3\mathcal{E}}\,g(\gamma)}
\end{eqnarray}
for $p_\pm$ orbitals, where
\begin{eqnarray}
\label{eq:hs}
h^s(\gamma)
&=&(1-\zeta_0)\sqrt{\frac{(1+\gamma^2)(1-\zeta_0^2)}{(\zeta_0^2/\gamma^2+1)(2\zeta_0^2/\gamma^2+\zeta_0^2+1)}}\\
\label{eq:hp0}
h^{p_0}(\gamma)
&=&h^{s}(\gamma)\,\frac{3\mathcal{E}}{2\mathcal{E}_0}
\left(1-\zeta_0\right)\sqrt{\frac{1+\gamma^2}{\zeta_0^2/\gamma^2+1}}\\
\label{eq:hp1}
h^{p_\pm}_\pm(\gamma)
&=&h^{s}(\gamma)\,\frac{3(1+\gamma^2)}{2(1-\zeta_0^2)}\left(\sqrt{\frac{\zeta_0^2/\gamma^2+1}{1+\gamma^2}}\mp\frac{\zeta_0}{\gamma}\,\mathrm{sgn}(m)\right)^2.
\end{eqnarray}
For $s$ orbitals, Eqs.\,(\ref{eq:final_g}) and (\ref{eq:hs}) coincide exactly with Eqs.\,(73) and (74) of Ref.\,\cite{PPT1}.
In the adiabatic limit $\gamma\ll1$, i.e.\ $\zeta_0(\gamma\ll1)\approx\gamma^2/3(1-28\gamma^2/45)$ (see Ref.\,\cite{PPT1} and Appendix 6), the exponent and prefactors in Taylor series up to second order in $\gamma$ are
\begin{eqnarray}
g(\gamma\ll1)&\approx& 1-\gamma^2/15\\
h^s(\gamma\ll1)&\approx&1\\
h^{p_0}(\gamma\ll1)&\approx&\frac{3\mathcal{E}}{2\mathcal{E}_0}\left(1+\frac{\gamma^2}{9}\right)\\
h^{p_\pm}_\pm(\gamma\ll1)&\approx&\frac32\mp\gamma\,\mathrm{sgn}(m)+\frac{\gamma^2}{3}.
\end{eqnarray}
Therefore, the ratio of the ionization rates for $p_\pm$ orbitals
\begin{eqnarray}
\frac{w^{p_-}_\pm(\mathcal{E},\omega)}
     {w^{p_+}_\pm(\mathcal{E},\omega)}
&\approx&
1\pm\frac{4\gamma}{3}+\frac{8\gamma^2}{9}
\qquad(\gamma\ll1)
\end{eqnarray}
is always larger than 1 for right circular polarization and smaller than 1 for left circular polarization.
That means, for e.g.\ right circular polarization, the ionization from $p_-$ orbitals is more preferred than the ionization from $p_+$ orbitals.
In the adiabatic case $\gamma=0$, i.e.\ the tunneling ist much faster than the rotation of the laser field, the ionization rates for $p_\pm$ orbitals are equal as expected.
In the non-adiabatic limit $\gamma\gg 1$, i.e.\ $\zeta_0(\gamma)\approx1-1/\ln\gamma$, the exponent and prefactors are approximated as
\begin{eqnarray}
g(\gamma\gg1)&\approx& \frac{3\ln\gamma}{2\gamma}\\
%h^s(\gamma\gg1)&\approx&\frac{\gamma}{(\ln\gamma)^{3/2}}\left(1-\frac{\mathcal{E}}{\mathcal{E}_0}\frac{1}{\gamma\ln\gamma}\right)\\
h^s(\gamma\gg1)&\approx&\frac{\gamma}{(\ln\gamma)^{3/2}}\\
h^{p_0}(\gamma\gg1)&\approx&\frac{3\mathcal{E}}{2\mathcal{E}_0}\frac{\gamma^2}{\left(\ln\gamma\right)^{5/2}}\\\nonumber
h^{p_\pm}_\pm(\gamma\gg1)
&=&\frac34\frac{\gamma}{(\ln\gamma)^{1/2}}\left[1\mp\left(1-\frac{1}{\ln\gamma}\right)\mathrm{sgn}(m)\right]^2.
\end{eqnarray}
Thus, the ratio of the ionization rates for $p_\pm$ orbitals is
\begin{eqnarray}
\frac{w^{p_-}_\pm(\mathcal{E},\omega)}
     {w^{p_+}_\pm(\mathcal{E},\omega)}
&\approx&
\left(2\ln\gamma\right)^{\pm2}\qquad(\gamma\gg1),
\end{eqnarray}
i.e.\ for right circular polarization, the ionization rate for $p_-$ orbitals is much larger than the one for $p_+$ orbitals.

\section{Results and discussion}

\begin{figure}[ht]
\includegraphics[width=\textwidth]{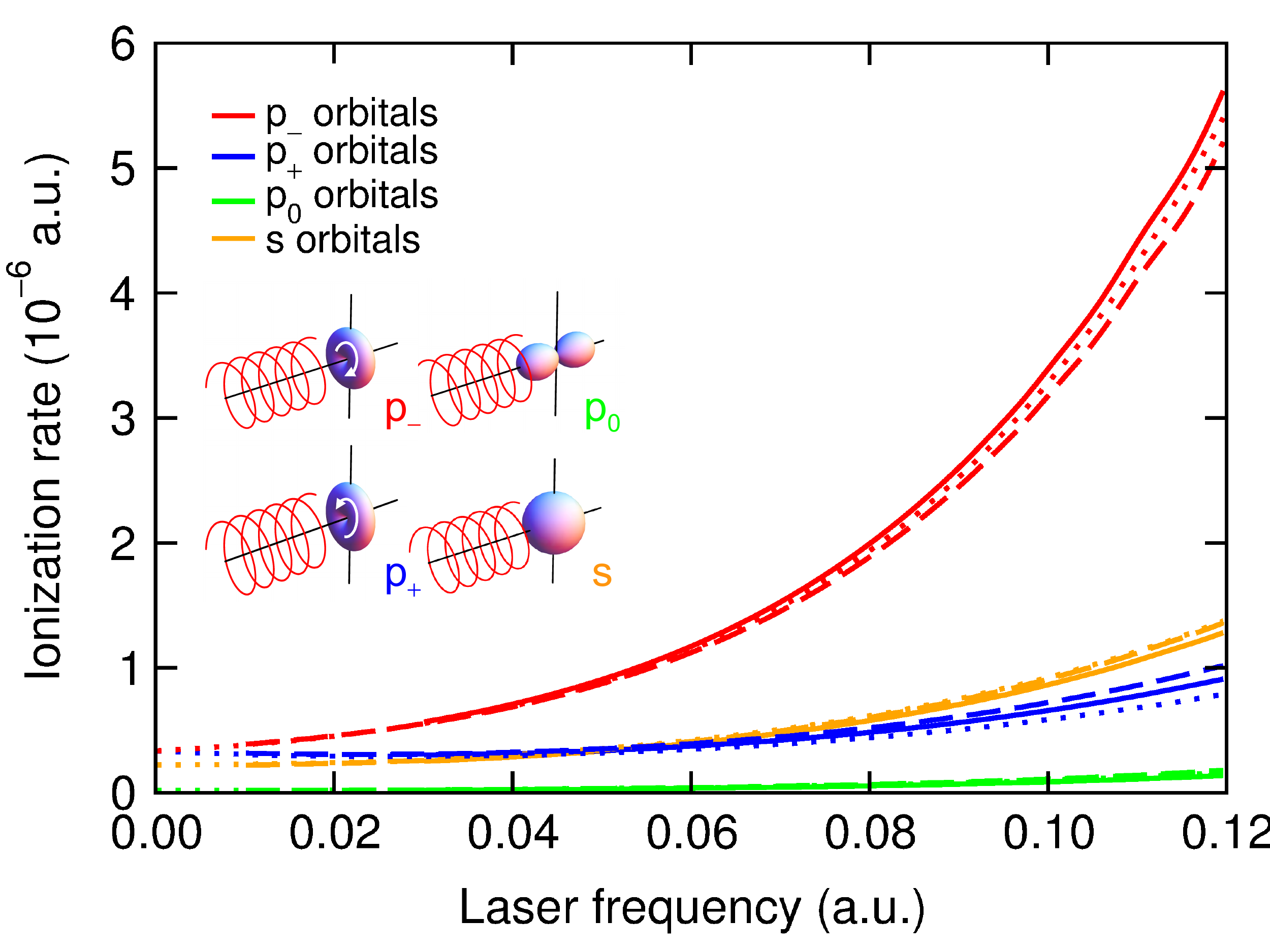}
\caption{}
\label{fig:sp}
\end{figure}

\begin{figure}
\includegraphics[width=\textwidth]{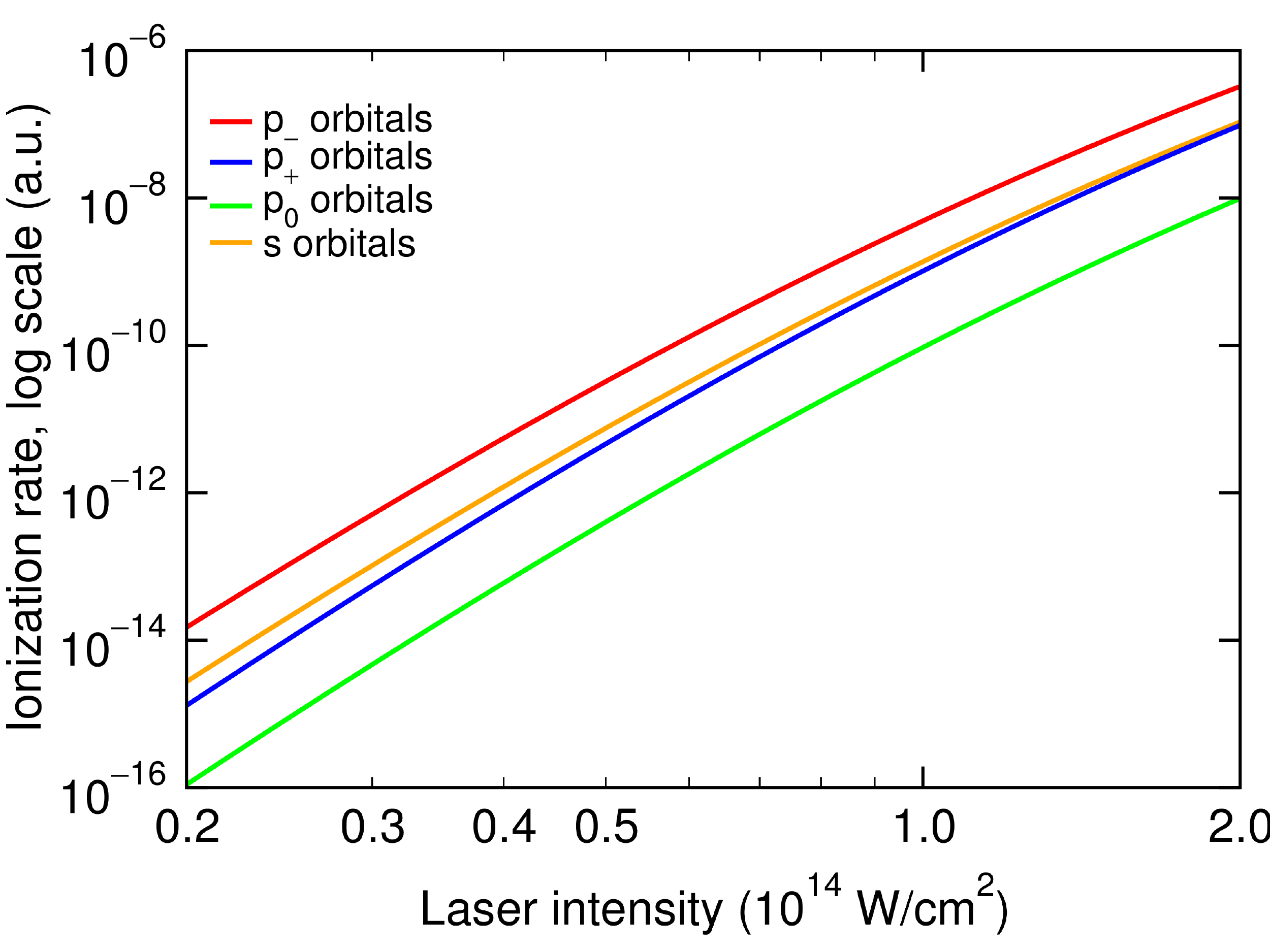}
\caption{}
\label{fig:intsp}
\end{figure}

\begin{figure}
\includegraphics[width=\textwidth]{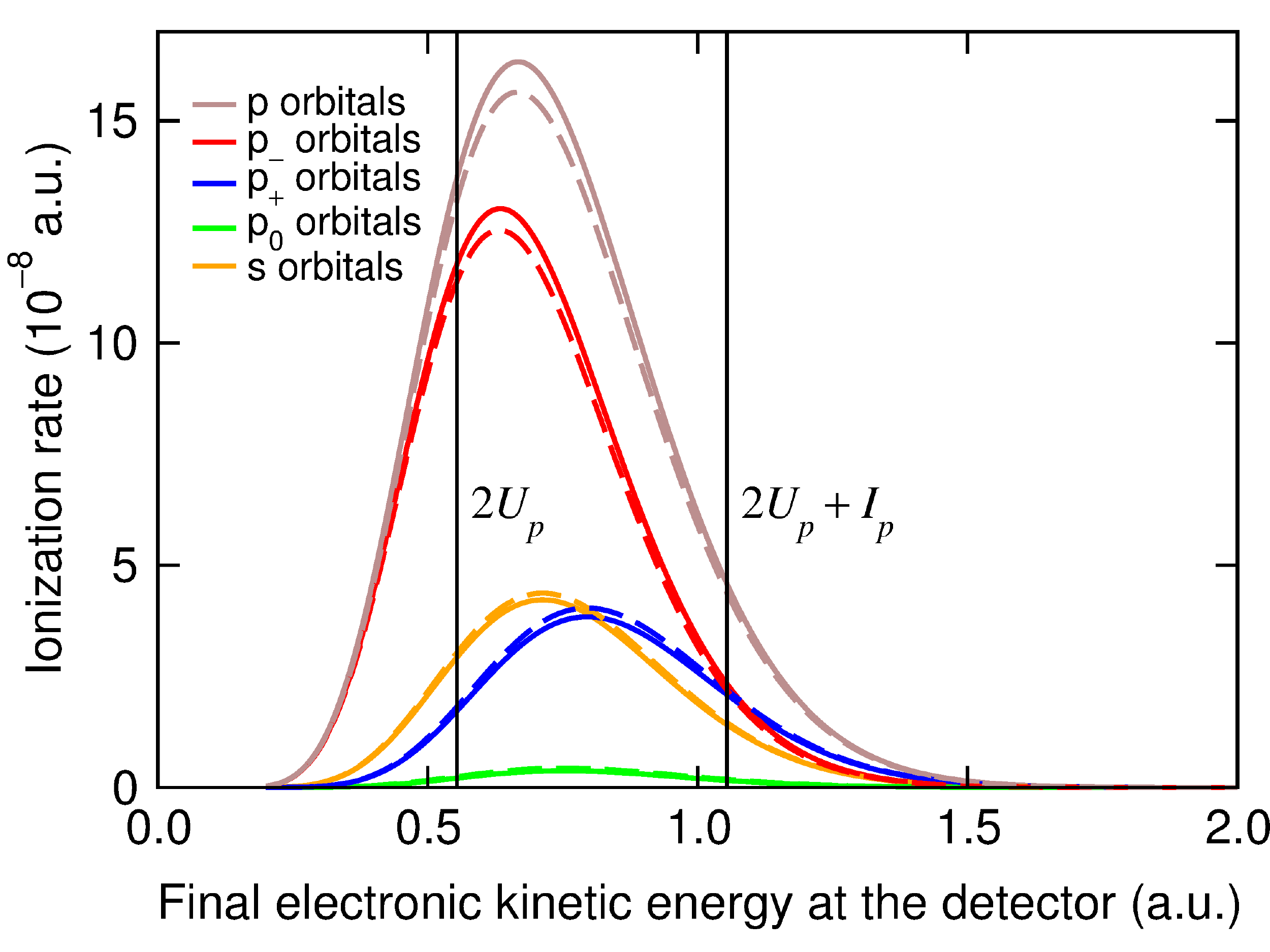}
\caption{}
\label{fig:spectsp}
\end{figure}

For application, we use Kr atom in the ground state with ionization potential $I_p=0.5$\,a.u.
An infrared circularly polarized strong laser field with typical experimental parameters of the laser frequency $\omega=0.057\,$a.u.\,(800\,nm) and
the laser amplitude $\mathcal{E}=0.06\,$a.u. ($I=2.5\cdot10^{14}$\,W/cm$^2$) ionizes an electron from the $4p$ valence orbital of the Kr atom.
In this experimental example, the Keldysh parameter is $\gamma=\omega/\mathcal{E}\sqrt{2I_p}=0.95$, thus an $4p$ electron tunnels the ionization barrier non-adiabatically with respect to the rotation of the electric field.
Although this is indeed non-adiabatic tunneling, many previous theoretical works are based on adiabatic approximation that cannot predict the difference of the ionization rates for $p_+$ and $p_-$ orbitals
in circularly polarized laser fields.
With our analytical formulas derived in Section II, which are beyond the original work \cite{PPT1,PPT2}, we present the results in Figs.\,\ref{fig:sp}--\ref{fig:spectsp} and show that the
ionization rates for $p_+$ and $p_-$ valence orbitals are indeed very different, supporting our physical interpretation in the previous work\,\cite{PRA1}.
We also show the results for the ionization rates for $p_0$ orbitals in Figs.\,\ref{fig:sp}--\ref{fig:spectsp}.
Although the results for the ionization rates for $s$ orbitals have nothing to do with the ionization of the Kr atom, we would like to include these results in Figs.\,\ref{fig:sp}--\ref{fig:spectsp} as well,
but these results cannnot be compared with the ones for $p$ orbitals due to the generally different factors $C_{\kappa l}$ for $s$ and $p$ orbitals depending on the model system.
However, in all calculations, we have used $C_{\kappa l}=1$ for simplicity.

In Fig.\,\ref{fig:sp}, the results for ionization rates depending on the laser frequency up to $\omega=0.12$\,a.u.\ for laser amplitude $\mathcal{E}=0.06\,$a.u.\ are shown.
The orange, green, blue, and red curves correspond to the rates for $s$, $p_0$, $p_+$, and $p_-$ orbitals, and
the associated solid, dashed, and dotted curves correspond to the accurate (Eqs.\,(\ref{eq:wsacc})--(\ref{eq:wp1acc})),
approximate (Eqs.\,(\ref{eq:wsappalt})--(\ref{eq:wp1appalt})), and simple (Eqs.\,(\ref{eq:wssimple})--(\ref{eq:wp1simple})) results, respectively.
For $s$ orbitals, the approximate and simple results coincide with each other very well within grafical resolution.
The accurate results for $s$ orbitals are a little separated from approximate and simple results, mainly due to the integral approximations (\ref{eq:int1}) and (\ref{eq:int2}) for $\omega\ll I_p$.
For $p_0$ orbitals, the ionization rates are very small compared to the ones for $p_\pm$ orbitals, supporting our thoughts in Section II.E, i.e.\
the destructive interference coming from two phase-opposite lobes of the $p_0$ oribtal causes ionization supression in the polarization plane perpendicular to the orbital nodal axis.
Again by more precise inspection, the approximate and simple results are similar
while the accurate results are a little separated from the approximate and simple results, again mainly due to approximations (\ref{eq:int1}) and (\ref{eq:int2}) for $\omega\ll I_p$.
While the ionization rates for $s$ and $p_0$ orbitals do not depend on the sense of circular polarization, the rates for $p_+$ and $p_-$ orbitals do.
For right circular polarization, the rates for $p_-$ orbitals are larger than the ones for $p_+$ orbitals by the ratio up to 6 for large frequencies.
For left circular polarization, the physical behaviour is reversed according to the fundamental symmetry in electrodynamics, i.e.\
the ionization rates for $p_\pm$ orbitals and left circular polarization are equal to the ones for $p_\mp$ orbitals and right circular polarization.
In the adiabatic limit $\gamma=\omega=0$, the rates for both $p_\pm$ orbitals are exactly equal as already predicted in many previous theoretical works based on adiabatic approximation.
For $p_\pm$ orbitals there are some (but not large) deviations between accurate, approximate, and simple results in particular for large frequencies or equivalently for large $\gamma$,
but in the adiabatic ($\gamma\ll1$) and non-adiabatic ($\gamma\sim1$) tunneling regimes, these three results for $p_+$ and for $p_-$ converge well.
The small differences for large frequencies are not only due to the saddle point method (which is only applicable for low frequencies)
but also due to the integral approximations (\ref{eq:int1}) and (\ref{eq:int2}) for $\omega\ll I_p$.

Fig.\,\ref{fig:intsp} shows the monotonically increased ionization rates for $s$ and $p$ orbitals versus laser intensity $I=c^2\varepsilon_0^2\mathcal{E}^2$ in the range from $2\cdot10^{13}$\,W/cm$^2$ to $2\cdot10^{14}$\,W/cm$^2$
(corresponding to the laser amplitude $\mathcal{E}$ in the range from $0.0169$\,a.u.\ to $0.0534$\,a.u.) for laser frequency $\omega=0.057\,$a.u.\,(800\,nm) in logarithmic scale.
In this figure, the accurate, approximate, and simple results coincide within graphical resolution.
As already explained above, the rates for $p_0$ orbitals are small compared to the rates for $p_\pm$ orbitals.
For right circular polarization, the rates for $p_-$ orbitals are larger than the ones for $p_+$ orbitals.
The corresponding ratio is large (small) for low (high) laser intensities or equivalently for large (small) $\gamma$.

Fig.\,\ref{fig:spectsp} shows the photoelectron spectra for $s$ and $p$ orbitals, right circular polarization, laser amplitude $\mathcal{E}=0.06\,$a.u., and laser frequency $\omega=0.057\,$a.u.\,(800\,nm).
The curves are calculated using time-averaged $n$-photon ionization rates $w_{n+}(\mathcal{E},\omega)$ (Eq.\,(\ref{eq:wn})) versus final electronic kinetic energy at the detector $k_n^2/2$ (Eq.\,(\ref{eq:conservation})).
Since there are no sums in simple results (Eqs.\,(\ref{eq:wssimple})--(\ref{eq:wp1simple})), only the corresponding accurate (solid) and approximate (dahsed) spectra
(cf.\,Eqs.\,(\ref{eq:wsappalt})--(\ref{eq:wp1appalt}) and Eqs.\,(\ref{eq:wssimple})--(\ref{eq:wp1simple}))
are presented in this figure, as well as the spectra for total $p$ orbitals according to $w_{n+}^p(\mathcal{E},\omega)=w_{n+}^{p_0}(\mathcal{E},\omega)+w_{n+}^{p_+}(\mathcal{E},\omega)+w_{n+}^{p_-}(\mathcal{E},\omega)$.
In fact, the spectra are different for electrons coming from different orbitals.
For right circular polarization, the ionization from $p_-$ orbitals is dominant, but there is a unique kinetic energy for which the $n$-photon ionization rates for $p_+$ and $p_-$ orbitals are equal.
This is the final kinetic energy $2U_p+I_p\approx1.05$\,a.u.\ for the approximate results.
For the accurate results, the intersection of the spectra for $p_+$ and $p_-$ orbitals lies at the energy a little more than $2U_p+I_p$.
Below (above) this unique electronic final kinetic energy, the ionization rates for $p_-$ orbitals are larger (smaller) than the ones for $p_+$ orbitals.
Therefore, the final kinetic energy indicates the strength and the direction of the ring current \cite{barthatom,barth} generated in the ion, measured in correlation with the electron.
Low energy electrons correlate to the ions with positive ring currents, while higher energy electrons correlate to the ions with negative ring currents.
Furthermore, the locations of the maxima for $s$, $p_0$, and total $p$ orbitals are similar, whereas the ones for $p_-$ orbitals are shifted to lower energy and the ones for $p_+$ orbitals are shifted to higher energy.
The reason is that the counter-clockwise (``positive'', right) sense of circular polarization drives the electron from $p_-$ and $p_+$ orbitals
with clockwise (``negative'') and counter-clockwise (``positive'') azimuthal velocities,
yielding smaller and larger kinetic energies, respectively.
In the adiabatic limit $\gamma\ll1$, all photoelectron spectra have its maxima at $2U_p\approx0.55\,$a.u.\ and in this case the spectra for $p_+$ and $p_-$ orbitals are identical.

\section{Conclusions}

We have validated the approximations used in our previous publication to derive simple formulas for ionization from different sub-states of the $p$ orbitals. We extended the PPT theory  to strong field non-adiabatic ioniziation for valence $p$ orbitals and we derived the corresponding ionization rates in full analytical form.
Due to the existence of the complex-valued tunneling angle in the prefactor of the ionization rate,
the rates are different for degenerate $p_+$ and $p_-$ orbitals and depend on the sense of rotation of the circularly polarized laser fields.
Strong field ionization preferentially removes a counter-rotating electron.
As expected ionization rates for degenerate $p_+$ and $p_-$ orbitals are significantly larger than the rates for $p_0$ orbitals due its orbital symmetry.
We have also demonstrated that ionization rates and electron spectra obtained in this approach are gauge-invariant, unlike the results of the strong field approximation.

An important extension of this work is the consideration of the electron spin \cite{BarthSO} to describe electronic ring currents in the ion, which couple electronic  spin and orbital degree of freedom.
Other possible extensions of this work include the theory of the non-adiabatic ionization for $p$ orbitals
in circularly or  elliptically polarized laser fields and static magnetic fields, see also works for $s$ orbitals \cite{rylyuk1,rylyuk}.

\acknowledgments{
We gratefully acknowledge stimulating discussions with E.\,Goulielmakis, M.\,Ivanov, U.\,Keller, J.\,Manz, and A.\,Wirth.
The work was supported by the DFG Grant No.\ Sm 292/2-1.}

\newpage
\section*{Appendix 1}

Here, we evaluate the $\phi$-integral in Eq.\,(\ref{eq:appendix1})
\begin{eqnarray}
\label{eq:I1}
I_1&=&\int_0^{2\pi}d\phi\,e^{i\rho((k_{2x}-k_{1x})\cos\phi+(k_{2y}-k_{1y})\sin\phi)}(f_\pm(k_{1\rho},\theta_1,\phi,t)+f_\pm(k_{2\rho},\theta_2,\phi,t)).
\end{eqnarray}
Using the expression for the function $f_\pm(k_\rho,\theta,\phi,t)$ (Eq.\,(\ref{eq:f})) and  the Euler's formula, the prefactor of the integrand in Eq.\,(\ref{eq:I1}) is rewritten as
\begin{eqnarray}
&&f_\pm(k_{1\rho},\theta_1,\phi,t)+f_\pm(k_{2\rho},\theta_2,\phi,t)\\\nonumber
&=&\frac12\left(k_{1\rho}e^{i\theta_1}+k_{2\rho}e^{i\theta_2}\pm 2iA_0e^{\pm i\omega t}\right)e^{-i\phi} +\frac12\left(k_{1\rho}e^{-i\theta_1}+k_{2\rho}e^{-i\theta_2}\mp 2iA_0e^{\mp i\omega t}\right)e^{i\phi},
\end{eqnarray}
hence
\begin{eqnarray}
I_1&=&\frac12\left(k_{1\rho}e^{i\theta_1}+k_{2\rho}e^{i\theta_2}\pm 2iA_0e^{\pm i\omega t}\right)\int_0^{2\pi}d\phi\,e^{-i\phi}e^{i\rho((k_{2x}-k_{1x})\cos\phi+(k_{2y}-k_{1y})\sin\phi)}\\\nonumber
&&+\frac12\left(k_{1\rho}e^{-i\theta_1}+k_{2\rho}e^{-i\theta_2}\mp 2iA_0e^{\mp i\omega t}\right)\int_0^{2\pi}d\phi\,e^{i\phi}e^{i\rho((k_{2x}-k_{1x})\cos\phi+(k_{2y}-k_{1y})\sin\phi)}.
\end{eqnarray}
We use the substitution $|\mathbf{k}_{2\parallel}-\mathbf{k}_{1\parallel}|\,\sin\phi'=(k_{2x}-k_{1x})\cos\phi+(k_{2y}-k_{1y})\sin\phi$, which is satisfied by the relations ($\sin\phi=\sqrt{1-\cos^2\phi}$)
\begin{eqnarray}
\cos\phi&=&\frac{(k_{2x}-k_{1x})\,\sin\phi'+(k_{2y}-k_{1y})\cos\phi'}{|\mathbf{k}_{2\parallel}-\mathbf{k}_{1\parallel}|}\\
\sin\phi&=&\frac{(k_{2y}-k_{1y})\sin\phi'-(k_{2x}-k_{1x})\,\cos\phi'}{|\mathbf{k}_{2\parallel}-\mathbf{k}_{1\parallel}|}.
\end{eqnarray}
With Euler's formula, we obtain
\begin{eqnarray}
e^{\pm i\phi}
&=&\frac{\mp i(k_{2x}-k_{1x})+(k_{2y}-k_{1y})}{|\mathbf{k}_{2\parallel}-\mathbf{k}_{1\parallel}|}\,e^{\pm i\phi'}\\
&=&\mp i\, \frac{k_{2\rho}e^{\pm i\theta_2}-k_{1\rho}e^{\pm i \theta_1}}{|\mathbf{k}_{2\parallel}-\mathbf{k}_{1\parallel}|}\,e^{\pm i\phi'}
\end{eqnarray}
and $d\phi=d\phi'$. The integral $I_1$ is then reexpressed as
\begin{eqnarray}
I_1&=&\frac{i}{2}\left(k_{1\rho}e^{i\theta_1}+k_{2\rho}e^{i\theta_2}\pm 2iA_0e^{\pm i\omega t}\right)\frac{k_{2\rho}e^{- i\theta_2}-k_{1\rho}e^{- i \theta_1}}{|\mathbf{k}_{2\parallel}-\mathbf{k}_{1\parallel}|}\int_0^{2\pi}d\phi'\,e^{-i\phi'}e^{i\rho|\mathbf{k}_{2\parallel}-\mathbf{k}_{1\parallel}|\,\sin\phi'}\qquad\\\nonumber
&&-\frac{i}{2}\left(k_{1\rho}e^{-i\theta_1}+k_{2\rho}e^{-i\theta_2}\mp 2iA_0e^{\mp i\omega t}\right)\frac{k_{2\rho}e^{ i\theta_2}-k_{1\rho}e^{ i \theta_1}}{|\mathbf{k}_{2\parallel}-\mathbf{k}_{1\parallel}|}\int_0^{2\pi}d\phi'\,e^{i\phi'}e^{i\rho|\mathbf{k}_{2\parallel}-\mathbf{k}_{1\parallel}|\,\sin\phi'}.
\end{eqnarray}
With the definition of the Bessel function of the first kind
\begin{eqnarray}
\label{eq:BesselJ}
J_n(x)&=&\frac{1}{2\pi}\int_0^{2\pi} d\phi\,e^{-in\phi}e^{ix\sin\phi}
=\frac{(-1)^n}{2\pi}\int_0^{2\pi} d\phi\,e^{in\phi}e^{ix\sin\phi},
\end{eqnarray}
the integral is evaluated as
\begin{eqnarray}
I_1&=&\pi i\left[\left(k_{1\rho}e^{i\theta_1}+k_{2\rho}e^{i\theta_2}\pm 2iA_0e^{\pm i\omega t}\right)\left(k_{2\rho}e^{- i\theta_2}-k_{1\rho}e^{- i \theta_1}\right)\right.\\\nonumber
&&\left.+\left(k_{1\rho}e^{-i\theta_1}+k_{2\rho}e^{-i\theta_2}\mp 2iA_0e^{\mp i\omega t}\right)\left(k_{2\rho}e^{ i\theta_2}-k_{1\rho}e^{ i \theta_1}\right)\right]
\frac{J_1\left(\rho|\mathbf{k}_{2\parallel}-\mathbf{k}_{1\parallel}|\right)}{{|\mathbf{k}_{2\parallel}-\mathbf{k}_{1\parallel}|}},
\end{eqnarray}
which is further simplified to
\begin{eqnarray}
I_1&=&2\pi i\left(k_{2\rho}^2-k_{1\rho}^2-2k_{2\rho} A_0\sin(\omega t\mp\theta_2)+2k_{1\rho} A_0\sin(\omega t\mp\theta_1)\right)\frac{J_1\left(\rho|\mathbf{k}_{2\parallel}-\mathbf{k}_{1\parallel}|\right)}{|\mathbf{k}_{2\parallel}-\mathbf{k}_{1\parallel}|},\qquad
\end{eqnarray}
cf.\ Eq.\,(\ref{eq:appendix1}).

\section*{Appendix 2}

In this appendix we prove the relation (Eq.\,\ref{eq:appendix2})
\begin{eqnarray}
\lim_{\rho\rightarrow\infty}\int d\mathbf{k}_\parallel\,g(\mathbf{k}_\parallel)\,\frac{\rho J_1(\rho k_\parallel)}{k_\parallel}&=&2\pi\int d\mathbf{k}_\parallel\,g(\mathbf{k}_\parallel)\delta(\mathbf{k}_\parallel)
\end{eqnarray}
with the arbitrary function $g(\mathbf{k}_\parallel)$ and $k_\parallel=k_\rho$ as
\begin{eqnarray}
\lim_{\rho\rightarrow\infty}\int d\mathbf{k}_\parallel\,g(\mathbf{k}_\parallel)\,\frac{\rho J_1(\rho k_\parallel)}{k_\parallel}
&=&\lim_{\rho\rightarrow\infty}\int_0^{2\pi} d\theta\int_0^\infty dk_\rho\,g(k_\rho\cos\theta\,\mathbf{e}_x+k_\rho\sin\theta\,\mathbf{e}_y)\rho J_1(\rho k_\rho)\qquad\\
&=&\lim_{\rho\rightarrow\infty}\int_0^{2\pi} d\theta\int_0^\infty dx\,g\left(\frac{x}{\rho}\,\cos\theta\,\mathbf{e}_x+\frac{x}{\rho}\,\sin\theta\,\mathbf{e}_y\right)J_1(x)\\
&=&\int_0^{2\pi} d\theta\int_0^\infty dx\,g(\mathbf{0})J_1(x)\\
&=&2\pi g(\mathbf{0})\int_0^{\infty} dx\,J_1(x)\\
&=&2\pi g(\mathbf{0})\\
&=&2\pi\int d\mathbf{k}_\parallel\,g(\mathbf{k}_\parallel)\delta(\mathbf{k}_\parallel).
\end{eqnarray}

\section*{Appendix 3}

In this section we will evaluate $k_+$- and $k_-$-integrals in Eq.\,(\ref{eq:jfluxlim}).
First with Eq.\,(\ref{eq:n0}), we rearrange the denominator of the $k_+$-integrand in Eq.\,(\ref{eq:jfluxlim}) as
\begin{eqnarray}
\nonumber
&&\left[\frac14(k_+-k_-)^2+k_y^2+k_z^2+A_0^2+2I_p-2n_1\omega+i\delta\right]\\\nonumber
&&\left[\frac14(k_++k_-)^2+k_y^2+k_z^2+A_0^2+2I_p-2n_2\omega- i\delta\right]\\
&=&\frac{1}{16}\left[(k_+-k_-)^2-4K_1+4i\delta\right]\left[(k_++k_-)^2-4K_2-4i\delta\right],
\end{eqnarray}
where $K_{1,2}=2(n_{1,2}-n_0)\omega-k_y^2-k_z^2$.
It has four zeros $k_+=k_{1\pm}, k_{2\pm}$, i.e.
\begin{eqnarray}
k_{1\pm}&=&\pm2\sqrt{K_1-i\delta}+k_-\\
k_{2\pm}&=&\pm2\sqrt{K_2+i\delta}-k_-.
\end{eqnarray}
Using Taylor series $\lim_{\delta\rightarrow 0}\sqrt{K\pm i\delta}=\lim_{\delta\rightarrow 0}\left(\sqrt{K}\pm i\delta/(2\sqrt{K})\right)$, the zeros are rewritten as
\begin{eqnarray}
k_{1\pm}&=&\pm2\sqrt{K_1}+k_-\mp\frac{i\delta}{\sqrt{K_1}}\\
k_{2\pm}&=&\pm2\sqrt{K_2}-k_-\pm\frac{i\delta}{\sqrt{K_2}}.
\end{eqnarray}
Thus, the $k_+$-integrand in Eq.\,(\ref{eq:jfluxlim}) is
\begin{eqnarray}
\tilde h_{\pm}(k_+)&=&
\left[\frac14(k_+-k_-)^2+k_y^2+k_z^2+A_0^2+2I_p-2n_1\omega+i\delta\right]^{-1}\\\nonumber
&&\left[\frac14(k_++k_-)^2+k_y^2+k_z^2+A_0^2+2I_p-2n_2\omega- i\delta\right]^{-1}h_\pm(k_+)\\
&=&\frac{16h_\pm(k_+)}{(k_+-k_{1+})(k_+-k_{1-})(k_+-k_{2+})(k_+-k_{2-})}.
\end{eqnarray}
The corresponding integral is then evaluated using the residue method as
\begin{eqnarray}
\label{eq:k+int}
\int_{-\infty}^\infty dk_+\,\tilde h_\pm(k_+)
&=&2\pi i\sum_{k=k_{1\pm},k_{2\pm}}\Theta(\mathrm{Im}\,k)\,\mathrm{Res}\,\tilde h_\pm(k),
\end{eqnarray}
where the residues are calculated as
\begin{eqnarray}
\mathrm{Res}\,\tilde h_\pm(k)&=&\lim_{k_+\rightarrow k}(k_+-k)\tilde h_\pm(k_+),
\end{eqnarray}
in particular
\begin{eqnarray}
\label{eq:res1}
\mathrm{Res}\,\tilde h_\pm(k_{1+})
%&=&\frac{16h_\pm(k_{1+})}{(k_{1+}-k_{1-})(k_{1+}-k_{2+})(k_{1+}-k_{2-})}\\
&=&\frac{8h_\pm(k_{1+})}{\left(2\sqrt{K_1}-\frac{i\delta}{\sqrt{K_1}}\right)}
\left(2\sqrt{K_1}-2\sqrt{K_2}+2k_--\frac{i\delta}{\sqrt{K_1}}-\frac{i\delta}{\sqrt{K_2}}\right)^{-1}\\\nonumber
&&\left(2\sqrt{K_1}+2\sqrt{K_2}+2k_--\frac{i\delta}{\sqrt{K_1}}+\frac{i\delta}{\sqrt{K_2}}\right)^{-1}\\
\label{eq:res2}
\mathrm{Res}\,\tilde h_\pm(k_{1-})
%&=&\frac{16h_\pm(k_{1-})}{(k_{1-}-k_{1+})(k_{1-}-k_{2+})(k_{1-}-k_{2-})}\\
&=&-\frac{8h_\pm(k_{1-})}{\left(2\sqrt{K_1}-\frac{i\delta}{\sqrt{K_1}}\right)}
\left(2\sqrt{K_1}+2\sqrt{K_2}-2k_--\frac{i\delta}{\sqrt{K_1}}+\frac{i\delta}{\sqrt{K_2}}\right)^{-1}\\\nonumber
&&\left(2\sqrt{K_1}-2\sqrt{K_2}-2k_--\frac{i\delta}{\sqrt{K_1}}-\frac{i\delta}{\sqrt{K_2}}\right)^{-1}\\
\label{eq:res3}
\mathrm{Res}\,\tilde h_\pm(k_{2+})
%&=&\frac{16h_\pm(k_{1+})}{(k_{1+}-k_{1-})(k_{1+}-k_{2+})(k_{1+}-k_{2-})}\\
&=&-\frac{8h_\pm(k_{2+})}{\left(2\sqrt{K_2}+\frac{i\delta}{\sqrt{K_2}}\right)}
\left(2\sqrt{K_1}-2\sqrt{K_2}+2k_--\frac{i\delta}{\sqrt{K_1}}-\frac{i\delta}{\sqrt{K_2}}\right)^{-1}\\\nonumber
&&\left(2\sqrt{K_1}+2\sqrt{K_2}-2k_--\frac{i\delta}{\sqrt{K_1}}+\frac{i\delta}{\sqrt{K_2}}\right)^{-1}\\
\label{eq:res4}
\mathrm{Res}\,\tilde h_\pm(k_{2-})
%&=&\frac{16h_\pm(k_{1+})}{(k_{1+}-k_{1-})(k_{1+}-k_{2+})(k_{1+}-k_{2-})}\\
&=&\frac{8h_\pm(k_{2-})}{\left(2\sqrt{K_2}+\frac{i\delta}{\sqrt{K_2}}\right)}
\left(2\sqrt{K_1}+2\sqrt{K_2}+2k_--\frac{i\delta}{\sqrt{K_1}}+\frac{i\delta}{\sqrt{K_2}}\right)^{-1}\\\nonumber
&&\left(2\sqrt{K_1}-2\sqrt{K_2}-2k_--\frac{i\delta}{\sqrt{K_1}}-\frac{i\delta}{\sqrt{K_2}}\right)^{-1}.
\end{eqnarray}
For $\delta=0$, $k_-=0$ and $0\neq n_1\neq n_2\neq 0$, i.e.\ $0\neq K_1\neq K_2\neq 0$, all four residues (\ref{eq:res1})--(\ref{eq:res4}) are finite.
Therefore, the $k_-$-integral in Eq.\,(\ref{eq:jfluxlim}) would be zero due to the existence of the factor $k_-$ in the integrand.
Thus, the condition $n=n_1=n_2$, i.e.\ $K=K_1=K_2$, must be satisfied and the residues (\ref{eq:res1})--(\ref{eq:res4}) are simplified to
\begin{eqnarray}
\label{eq:res1a}
\mathrm{Res}\,\tilde h_\pm(k_{1+})
&=&\frac{2h_\pm(k_{1+})}{\left(2\sqrt{K}-\frac{i\delta}{\sqrt{K}}\right)\left(2\sqrt{K}+k_-\right)\left(k_--\frac{i\delta}{\sqrt{K}}\right)}\\
\label{eq:res2a}
\mathrm{Res}\,\tilde h_\pm(k_{1-})
&=&\frac{2h_\pm(k_{1-})}{\left(2\sqrt{K}-\frac{i\delta}{\sqrt{K}}\right)\left(2\sqrt{K}-k_-\right)\left(k_-+\frac{i\delta}{\sqrt{K}}\right)}\\
\label{eq:res3a}
\mathrm{Res}\,\tilde h_\pm(k_{2+})
&=&-\frac{2h_\pm(k_{2+})}{\left(2\sqrt{K}+\frac{i\delta}{\sqrt{K}}\right)\left(2\sqrt{K}-k_-\right)\left(k_--\frac{i\delta}{\sqrt{K}}\right)}\\
\label{eq:res4a}
\mathrm{Res}\,\tilde h_\pm(k_{2-})
&=&-\frac{2h_\pm(k_{2-})}{\left(2\sqrt{K}+\frac{i\delta}{\sqrt{K}}\right)\left(2\sqrt{K}+k_-\right)\left(k_-+\frac{i\delta}{\sqrt{K}}\right)}.
\end{eqnarray}
For $K<0$, only two poles $k_{1+}$ and $k_{2+}$ have positive imaginary parts, therefore only two corresponding residues contribute to the $k_+$-integral (\ref{eq:k+int}),
but in the limit $\delta=0$ and $k_-=0$, the integrand of the $k_-$-integral in Eq.\,(\ref{eq:jfluxlim}) or the sum of these two residues times $k_-$ is exactly zero.
Therefore, we consider only the remaining case $K\geq 0$, where only two poles $k_{1-}$ and $k_{2+}$ have positive imaginary parts. Multiplying Eq.\,(\ref{eq:k+int}) by $k_-$, applying $\lim_{\delta\rightarrow 0}$,
and using residues (\ref{eq:res2a}) and (\ref{eq:res3a}) yields
\begin{eqnarray}
\lim_{\delta\rightarrow 0}k_-\int_{-\infty}^\infty dk_+\,\tilde h_\pm(k_+)
&=&2\pi i\,\frac{h_\pm(-2\sqrt{K}+k_-)-h_\pm(2\sqrt{K}-k_-)}{\sqrt{K}\left(2\sqrt{K}-k_-\right)}.
\end{eqnarray}
Inserting this result of the $k_+$-integral in Eq.\,(\ref{eq:jfluxlim}) and carrying the simple integration over $k_-$ yields the desired result ($n=n_1=n_2\geq n_0$ for $K\geq 0$)
\begin{eqnarray}
\lim_{\rho\rightarrow\infty}J_\pm(\rho,t)
&=&\pi\sum_{n\geq n_0}^\infty\int_{-\infty}^\infty dk_{y}\int_{-\infty}^\infty dk_z\,\frac{h_\pm(2\sqrt{K})-h_\pm(-2\sqrt{K})}{K},
\end{eqnarray}
where $K=2(n-n_0)\omega-k_y^2-k_z^2=k_n^2-k_y^2-k_z^2$, cf.\ Eq.\,(\ref{eq:jfluxlim1}).

\section*{Appendix 4}

Here, we derive the wavefunction in momentum representation $\tilde\varphi_{lm}(\mathbf{v}_\pm(t_i))$.
Using Fourier transformation (\ref{eq:phip}), wavefunction in coordinate representation $\varphi_{lm}(\mathbf{r})$ (\ref{eq:varphicoord}), spherical harmonics $Y_{lm}(\theta_r,\phi_r)$ (\ref{eq:harmonics}),
and equation for $-i\mathbf{v}_\pm(t_i)\mathbf{r}$ (\ref{eq:vt}), we obtain
\begin{eqnarray}
\tilde\varphi_{lm}(\mathbf{v}_\pm(t_i))&=&\frac{C_{\kappa l}\sqrt{\kappa}}{(2\pi)^{3/2}}\sqrt{\frac{2l+1}{4\pi}\frac{(l-m)!}{(l+m)!}}\int_0^\infty dr\,re^{-\kappa r}\\\nonumber
&&\int_0^\pi d\theta_r\,\sin\theta_r\,e^{-iv_\pm(t_i)r\cos\theta_{v\pm}(t_i)\cos\theta_r}\,P_l^m(\cos\theta_r)\\\nonumber
&&\int_0^{2\pi} d\phi_r\,e^{im\phi_r}e^{-iv_\pm(t_i)r\sin\theta_{v\pm}(t_i)\sin\theta_r\cos(\phi_{v\pm}(t_i)-\phi_r)}.
\end{eqnarray}
Using the substitution $\phi'=\phi_r-\phi_{v\pm}(t_i)-\pi/2$, we get $\cos(\phi_{v\pm}(t_i)-\phi_r)=-\sin\phi'$ and $d\phi_r=d\phi'$.
With the help of the Bessel function of the first kind (\ref{eq:BesselJ}), we evaluate the $\phi'$-integral as
\begin{eqnarray}
\tilde\varphi_{lm}(\mathbf{v}_\pm(t_i))&=&C_{\kappa l}\sqrt{\frac{\kappa}{2\pi}}\sqrt{\frac{2l+1}{4\pi}\frac{(l-m)!}{(l+m)!}}\,e^{im\phi_{v\pm}(t_i)}e^{-im\pi/2}
\int_0^\infty dr\,re^{-\kappa r}\\\nonumber
&&\int_0^\pi d\theta_r\,\sin\theta_r\,e^{-iv_\pm(t_i)r\cos\theta_{v\pm}(t_i)\cos\theta_r}\,P_l^m(\cos\theta_r)J_m(v_\pm(t_i)r\sin\theta_{v\pm}(t_i)\sin\theta_r).
\end{eqnarray}
Using Eqs.\,(19) and (23) of Ref.\ \cite{neves}, the $\theta_r$-integral is evaluated as
\begin{eqnarray}
\nonumber
&&\int_0^\pi d\theta_r\,\sin\theta_r\,e^{-iv_\pm(t_i)r\cos\theta_{v\pm}(t_i)\cos\theta_r}\,P_l^m(\cos\theta_r)J_m(v_\pm(t_i)r\sin\theta_{v\pm}(t_i)\sin\theta_r)\\
&=&\sqrt{\frac{2\pi}{v_\pm(t_i)r}}\,e^{-i(l-m)\pi/2}P_l^m(\cos\theta_{v\pm}(t_i))J_{l+1/2}(v_\pm(t_i)r),
\end{eqnarray}
hence
\begin{eqnarray}
\tilde\varphi_{lm}(\mathbf{v}_\pm(t_i))
&=&C_{\kappa l}\sqrt{\frac{\kappa}{v_\pm(t_i)}}\,Y_{lm}(\theta_{v\pm}(t_i),\phi_{v\pm}(t_i))\,e^{-il\pi/2}\int_0^\infty dr\,\sqrt{r}\,e^{-\kappa r}J_{l+1/2}(v_\pm(t_i)r),\qquad
\end{eqnarray}
cf.\ Eq.\,(\ref{eq:appendix4a}).
With the Taylor expression of the Bessel function of the first kind
\begin{eqnarray}
J_\alpha(x)&=&\sum_{\beta=0}^\infty\frac{(-1)^\beta}{\beta!\,\Gamma(\alpha+\beta+1)}\,\left(\frac{x}{2}\right)^{\alpha+2\beta},
\end{eqnarray}
we have
\begin{eqnarray}
\tilde\varphi_{lm}(\mathbf{v}_\pm(t_i))
&=&C_{\kappa l}\sqrt{\frac{\kappa}{v_\pm(t_i)}}\,Y_{lm}(\theta_{v\pm}(t_i),\phi_{v\pm}(t_i))\,e^{-il\pi/2}\\\nonumber
&&\sum_{\beta=0}^\infty\frac{(-1)^\beta}{\beta!\,\Gamma(\beta+l+3/2)}\left(\frac{v_\pm(t_i)}{2}\right)^{2\beta+l+1/2}\int_0^\infty dr\,r^{2\beta+l+1}\,e^{-\kappa r}.
\end{eqnarray}
Using the definition of the Gamma function ($z>0$)
\begin{eqnarray}
\Gamma(z)&=&\int_0^\infty dx\,x^{z-1}e^{-x},
\end{eqnarray}
the $r$-integration is easily carried out
\begin{eqnarray}
\int_0^\infty dr\,r^{2\beta+l+1}\,e^{-\kappa r}
%&=&\frac{1}{\kappa^{2\beta+l+2}}\int_0^\infty dx\,x^{2\beta+l+1}\,e^{-x}\\
&=&\frac{\Gamma(2\beta+l+2)}{\kappa^{2\beta+l+2}},
\end{eqnarray}
hence
\begin{eqnarray}
\tilde\varphi_{lm}(\mathbf{v}_\pm(t_i))
&=&\frac{C_{\kappa l}}{\sqrt{2\kappa^3}}\left(\frac{v_\pm(t_i)}{2\kappa}\right)^lY_{lm}(\theta_{v\pm}(t_i),\phi_{v\pm}(t_i))\,e^{-il\pi/2}\\\nonumber
&&\sum_{\beta=0}^\infty\frac{1}{\beta!}\frac{\Gamma(2\beta+l+2)}{\Gamma(\beta+l+3/2)}\left(-\frac{v_{\pm}(t_i)^2}{4\kappa^2}\right)^{\beta}.
\end{eqnarray}
Using the duplication formula for the Gamma function
\begin{eqnarray}
\label{eq:dupl}
\Gamma(2z)&=&\frac{\Gamma(z)\Gamma(z+1/2)}{2^{1-2z}\sqrt{\pi}}
\end{eqnarray}
for $z=\beta+l/2+1$ yields
\begin{eqnarray}
\label{eq:varphi1}
\tilde\varphi_{lm}(\mathbf{v}_\pm(t_i))
&=&C_{\kappa l}\,\sqrt{\frac{2}{\pi\kappa^3}}\left(\frac{v_\pm(t_i)}{\kappa}\right)^lY_{lm}(\theta_{v\pm}(t_i),\phi_{v\pm}(t_i))\,e^{-il\pi/2}\\\nonumber
&&\sum_{\beta=0}^\infty\frac{1}{\beta!}\frac{\Gamma(\beta+l/2+1)\Gamma(\beta+l/2+3/2)}{\Gamma(\beta+l+3/2)}\left(-\frac{v_\pm(t_i)^2}{\kappa^2}\right)^{\beta}.
\end{eqnarray}
Using the Gaussian hypergeometric series
\begin{eqnarray}
_2F_1(a,b;c;z)&=&\frac{\Gamma(c)}{\Gamma(a)\Gamma(b)}\sum_{\beta=0}^\infty\frac{\Gamma(\beta+a)\Gamma(\beta+b)}{\Gamma(\beta+c)}\frac{z^\beta}{\beta!}
\end{eqnarray}
and Eq.\,(\ref{eq:dupl}) for $z=l/2+1$, we obtain
\begin{eqnarray}
\tilde\varphi_{lm}(\mathbf{v}_\pm(t_i))
&=&\frac{C_{\kappa l}}{\sqrt{2\kappa^3}}\left(\frac{v_\pm(t_i)}{2\kappa}\right)^lY_{lm}(\theta_{v\pm}(t_i),\phi_{v\pm}(t_i))\,e^{-il\pi/2}\\\nonumber
&&\frac{\Gamma(l+2)}{\Gamma(l+3/2)}\,_2F_1\left(\frac{l}{2}+1,\frac{l}{2}+\frac32;l+\frac32;-\frac{v_\pm(t_i)^2}{\kappa^2}\right),
\end{eqnarray}
cf.\ Eq.\ (\ref{eq:varphiseries}).
Using Euler's hypergeometric transformation
\begin{eqnarray}
_2F_1(a,b;c;z)&=&(1-z)^{c-a-b}\,_2F_1(c-a,c-b;c;z),
\end{eqnarray}
the wavefunction is then rewritten as
\begin{eqnarray}
\label{eq:appendix4c}
\tilde\varphi_{lm}(\mathbf{v}_\pm(t_i))
&=&\frac{C_{\kappa l}}{v_\pm(t_i)^2+\kappa^2}\sqrt{\frac{2\kappa}{\pi}}\left(\frac{v_\pm(t_i)}{\kappa}\right)^lY_{lm}(\theta_{v\pm}(t_i),\phi_{v\pm}(t_i))\,e^{-il\pi/2}\\\nonumber
&&\frac{\sqrt{\pi}}{2^{l+1}}\frac{\Gamma(l+2)}{\Gamma(l+3/2)}\,_2F_1\left(\frac{l}{2}+\frac12,\frac{l}{2};l+\frac32;-\frac{v_\pm(t_i)^2}{\kappa^2}\right),
\end{eqnarray}
cf.\ Eq.\,(\ref{eq:varphifinal}).
With the Gaussian theorem for hypergeometric series
\begin{eqnarray}
_2F_1(a,b;c;1)&=&\frac{\Gamma(c)\Gamma(c-a-b)}{\Gamma(c-a)\Gamma(c-b)}
\end{eqnarray}
and Eq.\,(\ref{eq:dupl}) for $z=l/2+1$,
the series in Eq.\,(\ref{eq:appendix4c}) is convergent for $v_\pm(t_i)^2=-\kappa^2$, i.e.
\begin{eqnarray}
_2F_1\left(\frac{l}{2},\frac{l}{2}+\frac12;l+\frac32;1\right)&=&\frac{2^{l+1}}{\sqrt{\pi}}\,\frac{\Gamma(l+3/2)}{\Gamma(l+2)},
\end{eqnarray}
cf.\ Eq.\,(\ref{eq:2F1convergent}),
therefore the wavefunction in momentum representation (\ref{eq:appendix4c}) has the pole at $v_\pm(t_i)^2=-\kappa^2$.

\section*{Appendix 5}

Using Eqs.\,(\ref{eq:cosv}), (\ref{eq:sinv}), and $v_{\rho\pm}(t_i)^2|_{k=k_n}=-(k_z^2+2I_p)$, the factor $\left|e^{im\phi_{v\pm}(t_i)}\right|_{k=k_n}^2$ for $m=\pm 1$ is rewritten as
\begin{eqnarray}
\left|e^{im\phi_{v\pm}(t_i)}\right|_{k=k_n}^2
&=&\left|\cos\phi_{v\pm}(t_i)+i\,\mathrm{sgn}(m)\sin\phi_{v\pm}(t_i)\right|_{k=k_n}^2\\
&=&\frac{\left|v_{x\pm}(t_i)+i\,\mathrm{sgn}(m)v_{y\pm}(t_i)\right|_{k=k_n}^2}{k_z^2+2I_p}.
\end{eqnarray}
With Eqs.\,(\ref{eq:pxti}), (\ref{eq:pyti}), $k_\rho=\sqrt{k^2-k_z^2}$, and $\left|e^{im\theta}\right|^2=1$, it becomes
\begin{eqnarray}
\left|e^{im\phi_{v\pm}(t_i)}\right|_{k=k_n}^2
&=&\frac{\left[A_0\chi_n(k_z)\left(1\mp\mathrm{sgn}(m) \sqrt{1-1/\chi_n(k_z)^2}\right)-\sqrt{k_n^2-k_z^2}\right]^2}{k_z^2+2I_p}.
\end{eqnarray}
Using Eq.\,(\ref{eq:chinkz}), we obtain
\begin{eqnarray}
\left|e^{im\phi_{v\pm}(t_i)}\right|_{k=k_n}^2
&=&\frac{\left[A_0^2\chi_n(k_z)^2\left(1\mp\mathrm{sgn}(m) \sqrt{1-1/\chi_n(k_z)^2}\right)-n\omega\right]^2}{A_0^2\chi_n(k_z)^2\left(k_z^2+2I_p\right)}.
\end{eqnarray}
Using $\gamma^2=2I_p/A_0^2$ and Eq.\,(\ref{eq:n0}), i.e.\ $n_0\omega=A_0^2/2+I_p=I_p/\gamma^2+I_p=I_p(1+\gamma^2)/\gamma^2$, we obtain the result for $m=\pm 1$
\begin{eqnarray}
\left|e^{im\phi_{v\pm}(t_i)}\right|_{k=k_n}^2
&=&\frac{I_p\left[2\chi_n(k_z)^2\left(1\mp\,\mathrm{sgn}(m)\sqrt{1-1/\chi_n(k_z)^2}\right)-(1+\gamma^2)n/n_0\right]^2}{2\gamma^2\chi_n(k_z)^2\left(k_z^2+2I_p\right)},
\end{eqnarray}
cf.\ Eq.\,(\ref{eq:expim}).

\section*{Appendix 6}

To obtain the simple analytic solution $\zeta_0(\gamma)$ of Eq.\,(\ref{eq:eqt0g})
\begin{eqnarray}
\mathrm{artanh}\sqrt{\frac{\zeta_0^2+\gamma^2}{1+\gamma^2}}&=&\frac{1}{1-\zeta_0}\sqrt{\frac{\zeta_0^2+\gamma^2}{1+\gamma^2}},
\end{eqnarray}
in the adiabatic limit $\gamma\ll 1$, we use power series
\begin{eqnarray}
\zeta_0(\gamma)&=&\sum_{i=0}^\infty c_i\gamma^i,
\end{eqnarray}
yielding
\begin{eqnarray}
\label{eq:eq1}
\mathrm{artanh}\sqrt{\frac{\left(\sum_{i=0}^\infty c_i\gamma^i\right)^2+\gamma^2}{1+\gamma^2}}-\frac{1}{1-\sum_{i=0}^\infty c_i\gamma^i}\sqrt{\frac{\left(\sum_{i=0}^\infty c_i\gamma^i\right)^2+\gamma^2}{1+\gamma^2}}&=&0.
\end{eqnarray}
In the Taylor series of Eq.\,(\ref{eq:eq1}) at $\gamma=0$, each coefficient of powers $\gamma^i$ must be zero. For zeroth order, we get
\begin{eqnarray}
\mathrm{artanh}\,c_0-\frac{c_0}{1-c_0}&=&0,
\end{eqnarray}
with the solution $c_0=0$. The coefficient for the first order is then automatically zero. For the second order, we get
\begin{eqnarray}
 \frac{c_1(1+c_1^2)}{\sqrt{1+c_1^2}}&=&0,
\end{eqnarray}
thus $c_1=0$.
For other orders, we get
\begin{eqnarray}
\frac{1}{3}-c_2&=&0\\
c_3&=&0\\
\frac{28}{135}+c_4&=&0,
\end{eqnarray}
thus $c_2=1/3$, $c_3=0$, and $c_4=-28/135$.
The solution of Eq.\ (\ref{eq:eqt0g}) is therefore
\begin{eqnarray}
\label{eq:t0g}
\zeta_0(\gamma)&=&\frac{\gamma^2}{3}\left(1-\frac{28}{45}\,\gamma^2+\frac{236}{567}\,\gamma^4-\frac{5212}{18225}\,\gamma^6+\frac{12570692}{63149625}\,\gamma^8-\cdots\right)
\end{eqnarray}
and for small $\gamma\ll 1$
\begin{eqnarray}
\label{eq:t0gapprox}
\zeta_0(\gamma)&\approx&\frac{\gamma^2}{3}.
\end{eqnarray}

\newpage
\section*{Figure Captions}

\noindent
FIG. 1: Time-averaged ionization rates $w_+(\mathcal{E},\omega)$ for $s$ (orange), $p_0$ (green), $p_+$ (blue), $p_-$ (red) orbitals and right circular polarization versus laser frequency $\omega$ for $I_p=0.5\,$a.u.\ and
laser amplitude $\mathcal{E}=0.06\,$a.u.
The solid, dashed, and dotted curves correspond to the accurate (Eqs.\,(\ref{eq:wsacc})--(\ref{eq:wp1acc})), approximate (Eqs.\,(\ref{eq:wsappalt})--(\ref{eq:wp1appalt})),
and simple (Eqs.\,(\ref{eq:wssimple})--(\ref{eq:wp1simple})) results, respectively.
Note that $|C_{\kappa l}|^2=1$ are used.

\vspace{1cm}
\noindent
FIG. 2: Time-averaged ionization rates $w_+(\mathcal{E},\omega)$ for $s$ (orange), $p_0$ (green), $p_+$ (blue), $p_-$ (red) orbitals and right circular polarization
versus laser intensity $I=c^2\varepsilon_0^2\mathcal{E}^2$ for $I_p=0.5\,$a.u.\ and laser frequency $\omega=0.057\,$a.u.\,(800\,nm) in logarithmic scale.
The solid curves corresponds to the accurate results (Eqs.\,(\ref{eq:wsacc})--(\ref{eq:wp1acc})).
Because of the logarithmic scale, the accurate (Eqs.\,(\ref{eq:wsacc})--(\ref{eq:wp1acc})), approximate (Eqs.\,(\ref{eq:wsappalt})--(\ref{eq:wp1appalt})),
and simple (Eqs.\,(\ref{eq:wssimple})--(\ref{eq:wp1simple})) results coincide within graphical resolution.
Note that $|C_{\kappa l}|^2=1$ are used.

\vspace{1cm}
\noindent
FIG. 3: Time-averaged $n$-photon ionization rates $w_{n+}(\mathcal{E},\omega)$ (Eq.\,(\ref{eq:wn})) or equivalently photoelectron energy distribution at the detector
for $s$ (orange), $p_0$ (green), $p_+$ (blue), $p_-$ (red), and total $p$ (brown) orbitals and right circular polarization
versus final electronic kinetic energy $k_n^2/2$ (Eq.\,(\ref{eq:conservation})) for $I_p=0.5\,$a.u., laser amplitude $\mathcal{E}=0.06\,$a.u., and laser frequency $\omega=0.057\,$a.u.\,(800\,nm).
The solid and dashed curves correspond to the accurate (Eqs.\,(\ref{eq:wsacc})--(\ref{eq:wp1acc})) and approximate (Eqs.\,(\ref{eq:wsappalt})--(\ref{eq:wp1appalt})) results, respectively.
The spectra for total $p$ orbitals are calculated according to $w_{n+}^p(\mathcal{E},\omega)=w_{n+}^{p_0}(\mathcal{E},\omega)+w_{n+}^{p_+}(\mathcal{E},\omega)+w_{n+}^{p_-}(\mathcal{E},\omega)$.
Note that $|C_{\kappa l}|^2=1$ are used.
The approximate results of the ionization rates for $p_+$ and $p_-$ orbitals are equal at the final kinetic energy $2U_p+I_p\approx 1.05\,$a.u, see text for discussion.
In the adiabatic limit $\gamma\ll1$, all photoelectron distributions are peaked at $2U_p\approx 0.55\,$a.u.\ and are the same for $p_+$ and $p_-$ orbitals.

\newpage
\section*{Figures}

\newpage
\begin{thebibliography}{999}

\bibitem{keldysh} L. V. Keldysh, Sov. Phys. JETP \textbf{20}, 1307  (1965).
\bibitem{ursi1} P. Eckle, M. Smolarski, P. Schlup, J. Biegert, A. Staudte, M. Sch\"offler, H. G. Muller, R. D\"orner, and U. Keller, Nature Phys. \textbf{4}, 565 (2008).
\bibitem{ursi2} P. Eckle, A. N. Pfeiffer, C. Cirelli, A. Staudte, R. D\"orner, H. G. Muller, M. B\"uttiker, and U. Keller, Science \textbf{322} 1525, (2008).
\bibitem{corkum1} H. Akagi, T. Otobe, A. Staudte, A. Shiner, F. Turner, R. D\"orner, D. M. Villeneuve, and P. B. Corkum, Science \textbf{325}, 1364 (2009).
\bibitem{torres} R. Torres, T. Siegel, L. Brugnera, I. Procino, J. G. Underwood, C. Altucci, R. Velotta, E. Springate, C. Froud, I. C. E. Turcu, M. Yu. Ivanov, O. Smirnova, and J. P. Marangos, Opt. Express \textbf{18}, 3174 (2010).
\bibitem{ursi3} A. P. Pfeiffer, C. Cirelli, M. Smolarski, R. D\"orner, and U. Keller, Nature Phys. \textbf{7}, 428 (2011).
\bibitem{ursi4} A. P. Pfeiffer, C. Cirelli, M. Smolarski, D. Dimitrovski, M. Abu-samha, L. B. Madsen, and U. Keller, Nature Phys. \textbf{8}, 76 (2012).
\bibitem{Anatomy} M. Yu. Ivanov, M. Spanner, and O. Smirnova, J. Mod. Opt. \textbf{52}, 165 (2005).
\bibitem{YudinIvanov} G. L. Yudin and M. Yu. Ivanov, Phys. Rev. A \textbf{64}, 013409 (2001).
\bibitem{PRA1} I. Barth and O. Smirnova, Phys. Rev. A 84, 063415 (2011), Errata: Phys. Rev. A 85, 029906(E) (2012), Phys. Rev. A \textbf{85}, 039903(E) (2012).
\bibitem{herath} T. Herath, L. Yan, S. K. Lee, and W. Li, Phys. Rev. Lett. \textbf{109}, 043004 (2012).
%\bibitem{bethe} H. Bethe, \textit{Intermediate Quantum Mechanics} (W. A. Benjamin, Inc., New York, 1964).
%\bibitem{Rzazewski1} K. Rzazewski and B. Piraux, Phys. Rev. A \textbf{47}, R1612 (1993).
%\bibitem{Rzazewski2} J. Zakrzewski, D. Delande, J. C. Gay, and K. Rzazewski, Phys. Rev. A \textbf{47}, R2468 (1993).
\bibitem{PPT1} A. M. Perelomov, V. S. Popov, and M. V. Terent'ev, Sov. Phys. JETP \textbf{23}, 924 (1966).
\bibitem{PPT2} A. M. Perelomov, V. S. Popov, and M. V. Terent'ev, Sov. Phys. JETP \textbf{24}, 207 (1967).
\bibitem{PPT3} A. M. Perelomov and V. S. Popov, Sov. Phys. JETP \textbf{25}, 336 (1967).
\bibitem{PPTnew} S. V. Popruzhenko, V. D. Mur, V. S. Popov, and D. Bauer, Phys. Rev. Lett. \textbf{101}, 193003 (2008).
\bibitem{Lisa} L. Torlina and O. Smirnova, Phys. Rev. A \textbf{86}, 043408 (2012).
\bibitem{Jivesh} J. Kaushal and O. Smirnova, Phys. Rev. A, submitted.
\bibitem{jphysb} E. V. Koryukina, J. Phys. B \textbf{38}, 3296 (2005).
\bibitem{frolov1} M. V. Frolov, N. L. Manakov, E. A. Pronin, and A. F. Starace, Phys. Rev. Lett. \textbf{91}, 053003 (2003).
\bibitem{frolov2} M. V. Frolov, N. L. Manakov, E. A. Pronin, and A. F. Starace, J. Phys. B \textbf{36}, L419 (2003).
\bibitem{DBauer} D. Bauer, D. B. Milo\v{s}evi\'c, and W. Becker, Phys. Rev. A \textbf{72}, 023415 (2005).
\bibitem{Smirnovajmo07} O. Smirnova, M. Spanner, and M. Ivanov, J. Mod. Opt. \textbf{54}, 1019 (2007).
\bibitem{bauer} J. H. Bauer, Phys. Rev. A \textbf{83}, 035402 (2011).
\bibitem{bauer2} J. H. Bauer, Phys. Rev. A \textbf{84}, 025403 (2011).
\bibitem{Gribakin}G. F. Gribakin and M. Y. Kuchiev, Phys. Rev. A \textbf{55}, 3760 (1997).
\bibitem{barthatom} I. Barth and J. Manz, Phys. Rev. A \textbf{75}, 012510 (2007).
\bibitem{barth} I. Barth and J. Manz, in \textit{Progress in Ultrafast Intense Laser Science VI}, edited by K. Yamanouchi, A.D. Bandrauk, and G. Gerber, Springer Series in Chemical Physics Vol. 99 (Springer, Berlin, 2010).
\bibitem{goul10} E. Goulielmakis, Z.-H. Loh, A. Wirth, R. Santra, N. Rohringer, V. S. Yakovlev, S. Zherebtsov, T. Pfeifer, A. M. Azzeer, M. F. Kling, S. R. Leone, and F. Krausz, Nature \textbf{466}, 739 (2010).
\bibitem{neves} A. A. R. Neves, L. A. Padilha, A. Fontes, E. Rodriguez, C. H. B. Cruz, L. C. Barbosa, and C. L. Cesar, J. Phys. A \textbf{39}, L293 (2006).
%\bibitem{yudin_ivanov} ??? G. Yudin and M. Yu. Ivanov, Phys. Rev. A \textbf{64}, 013409 (2001).
\bibitem{BarthSO} I. Barth and O. Smirnova, in preparation.
\bibitem{rylyuk1} V. M. Rylyuk and J. Ortner, Phys. Rev. A \textbf{67}, 013414 (2003).
\bibitem{rylyuk} V. M. Rylyuk, Phys. Rev. A \textbf{86}, 013402 (2012).

\end{thebibliography}
\end{document}